\documentclass[%
 reprint,
superscriptaddress,
 amsmath,amssymb,
 aps,onecolumn
]{revtex4-2}

\usepackage[english]{babel}

\usepackage{amsmath}
\usepackage{graphicx}
\usepackage{setspace}
\usepackage{lineno}
\usepackage{xfrac}
\usepackage{float}
\usepackage{multirow}
\usepackage{mathrsfs}
\usepackage{upgreek}
\usepackage{lineno}
\usepackage[colorlinks=true, citecolor=black, linkcolor=black, urlcolor=black]{hyperref}
\usepackage{color, colortbl}
\usepackage{lettrine}
\usepackage{gensymb}

\usepackage[colorlinks=true, citecolor=black, linkcolor=black, urlcolor=black]{hyperref}

\definecolor{Gray}{gray}{0.9}

\usepackage{xcolor}
\usepackage{soulutf8}

\usepackage[most]{tcolorbox}
\newtcolorbox{highlighted}{colback=yellow,coltext=black,breakable}

\spacing{1.25}

\begin{document}

\title{Dynamic Imaging of Periodic Structures using Extreme Ultraviolet Scatterometry}

\author{Brendan McBennett}
\affiliation{Department of Physics, JILA, and STROBE NSF Science and Technology Center, University of Colorado and NIST, Boulder, Colorado 80309 USA}
\affiliation{Applied Physics Division, National Institute of Standards and Technology, Boulder, Colorado 80305, USA}
\author{Michael Tanksalvala}
\affiliation{Applied Physics Division, National Institute of Standards and Technology, Boulder, Colorado 80305, USA}
\author{Emma E. Nelson}
\affiliation{Department of Physics, JILA, and STROBE NSF Science and Technology Center, University of Colorado and NIST, Boulder, Colorado 80309 USA}
\author{Theodore H. Culman}
\affiliation{Department of Physics, JILA, and STROBE NSF Science and Technology Center, University of Colorado and NIST, Boulder, Colorado 80309 USA}
\author{Yunhao Li}
\affiliation{Department of Physics, JILA, and STROBE NSF Science and Technology Center, University of Colorado and NIST, Boulder, Colorado 80309 USA}
\author{Jiayi Liu}
\affiliation{Department of Physics, JILA, and STROBE NSF Science and Technology Center, University of Colorado and NIST, Boulder, Colorado 80309 USA}
\author{Ethan Berk}
\affiliation{Department of Physics, JILA, and STROBE NSF Science and Technology Center, University of Colorado and NIST, Boulder, Colorado 80309 USA}
\author{Albert Beardo}
\affiliation{Department of Physics, Universitat Autònoma de Barcelona, 08193 Bellaterra, Catalonia, Spain}
\author{James Harford}
\affiliation{Physics Department, Utah State University, Logan, Utah, 84322 USA}
\author{Justin Shaw}
\affiliation{Applied Physics Division, National Institute of Standards and Technology, Boulder, Colorado 80305, USA}
\author{Henry C. Kapteyn}
\affiliation{Department of Physics, JILA, and STROBE NSF Science and Technology Center, University of Colorado and NIST, Boulder, Colorado 80309 USA}
\affiliation{KMLabs Inc., 4775 Walnut St. \#102, Boulder, Colorado 80309 USA}
\author{Margaret M. Murnane}
\affiliation{Department of Physics, JILA, and STROBE NSF Science and Technology Center, University of Colorado and NIST, Boulder, Colorado 80309 USA}
\author{Joshua L. Knobloch}
\affiliation{Physics Department, Utah State University, Logan, Utah, 84322 USA}

\maketitle

\textbf{ABSTRACT:}
Dynamic scattering and imaging with coherent, ultrafast, extreme ultraviolet (EUV) light sources can resolve charge, phonon and spin processes on their intrinsic length and time scales. However, full field coherent diffraction imaging requires scanning of the sample combined with computational phase retrieval, making it challenging to quickly acquire a large series of dynamic frames. In this work, we demonstrate a technique for extracting dynamic 1D images of the average unit cell in a periodic sample from traditional EUV scatterometry data by analyzing the changing intensities of the far field diffracted orders. Starting from a system of equations relating small changes in far field diffraction to phase and amplitude perturbations at the sample plane, it is shown that under certain conditions, changes to the $n$th diffracted order map exclusively onto the $n$th Fourier component of the perturbation via a closed-form relation. We show through rigorous coupled-wave analysis simulations that our method can provide a good approximation even outside the scalar diffraction theory framework in which it is derived. Finally, we experimentally demonstrate this reconstruction method by exiting 1D nickel nanowires on a diamond substrate using an infrared laser pump pulse, and measuring their relaxation using a time-delayed EUV probe pulse, to visualize nanoscale phonon dynamics.

\section{Introduction}
\label{sec:one}

A major metrology and materials physics challenge is to develop laboratory scale techniques capable of imaging charge, spin and thermal dynamics on their intrinsic length and time scales. Scanning probe microscopy and transmission electron microscopy offer exquisite spatial resolution, but face challenges in integrating ultrafast pumps for dynamic metrology \cite{Liang2023,Alcorn2023,Jing2019,Zhao2025}, while ultrafast lasers provide femtosecond temporal resolution but have limited spatial resolution at visible wavelengths \cite{Delor2020,Veysset2017}. In recent years, tabletop approaches to generating coherent extreme ultraviolet (EUV) and soft x-ray light based on high harmonic generation have greatly expanded the ability to probe nanoscale dynamic processes \cite{Rundquist1998}. Moreover, by combining coherent EUV beams with coherent diffraction imaging and phase retrieval algorithms \cite{Tanksalvala2021}, the limits of imperfect EUV focusing optics can be overcome. For instance, ptychography can be used to image near the diffraction limit \cite{Gardner2017,Eschen2025,Jiang2018}, to image with temporal resolution matching the source \cite{Karl2018,Zayko2021}, and can even provide three-dimensional reconstructions of nanoscale devices \cite{Tanksalvala2021,Holler2017}.  Most computational techniques for imaging with EUV light are data-intensive, thus imposing increased data acquisition and processing times.  As a result, advances are still being made to improve the speed of these methods, by altering the algorithms or techniques \cite{Kang2023}. A yet-unmet capability, toward the goal of imaging nanoscale transport at its intrinsic length and time scales, would be a technique that simultaneously accesses near-wavelength spatial information, can be implemented stroboscopically, and furthermore provides quantifiable error bars.

Fortunately, diffraction measurements can provide spatial information about a nanostructured sample even without a full image reconstruction. For example, transient grating experiments track the evolution of a sinusoidal excitation generated through the interference of two pump beams via the diffraction of a time-delayed probe \cite{Bencivenga2023}; however, it is difficult to combine this approach with nanostructured samples where multiple spatial modes may be excited simultaneously. Other diffraction-based techniques can speed up data acquisition significantly but may still require computationally-intensive, iterative reconstruction of the sample structure. Scatterometry, where coherent light of variable wavelength, angle or polarization diffracts from a surface profile, provides a rapid means of defect recognition and process development feedback in industrial-scale nanofabrication \cite{Madsen2016}. EUV light is often employed for scatterometry because of its short wavelength and because tunable EUV sources near atomic resonances provide element-resolved information \cite{Ku2016, Wang2021, Tanksalvala2021, Esashi2023}. The approach to interpreting scatterometry data is model-based, wherein simulated diffraction profiles are compared to observations until a fit is achieved; it can be greatly simplified when certain details about the sample geometry and composition are already known. Scatterometry experiments can also measure dynamics in periodic systems using a time delayed probe pulse which measures the change in diffraction efficiency after an ultrafast pump excitation. Unlike transient reflectivity experiments which measure a change in real refractive index, diffraction from a periodic profile also captures surface deformation due to the phase difference between light reflected from the structure and substrate \cite{Tobey2007}. Prior dynamic EUV scatterometry experiments have investigated heat flow from metallic nanoscale gratings down to $\sim$50~nm period on dielectric substrates using off-resonance EUV light which is sensitive to sub-angstrom thermal expansion rather than electron dynamics. However, these studies aggregate the change in diffraction efficiency into a single time-dependent scalar, which discards spatial information present in the diffraction profile \cite{Frazer2019}.

In this work, we demonstrate how EUV scatterometry data can be used for dynamic imaging of periodic structures (DIPS) in one dimension, to spatially resolve the dynamic perturbation to a known average unit cell. In section \ref{sec:two}, using scalar Fresnel diffraction theory, we derive a linear relationship between the change in intensity of each diffracted order and the corresponding Fourier components of the dynamic perturbation. For even phase perturbations to certain static profiles, the problem simplifies and it is only necessary to know the perturbation in the $n$th diffracted order to solve for the $n$th component of the sample perturbation, as confirmed by numerical simulations in section \ref{sec:three}. Section \ref{sec:four} then discusses how to incorporate the effect of amplitude perturbations, since in most scatterometry experiments the dynamics which produce a phase perturbation will also induce a change in reflectivity. Rigorous coupled-wave analysis simulations in section \ref{sec:five} establish that the approach can provide useful results even outside the initial scalar Fresnel diffraction theory approximations. Finally in section \ref{sec:experiment}, we use EUV scatterometry data to reconstruct the surface deformation of a 1D square nickel grating on a diamond substrate after excitation by an ultrafast infrared pump and compare with finite element predictions. The ability of scatterometry experiments to provide spatially-resolved information about dynamics in periodic systems allows for a new rapid metrology approach alongside traditional dynamic imaging techniques.

\section{Reconstruction method}
\label{sec:two}

\begin{figure}
    \centering
    \includegraphics[width=1\linewidth]{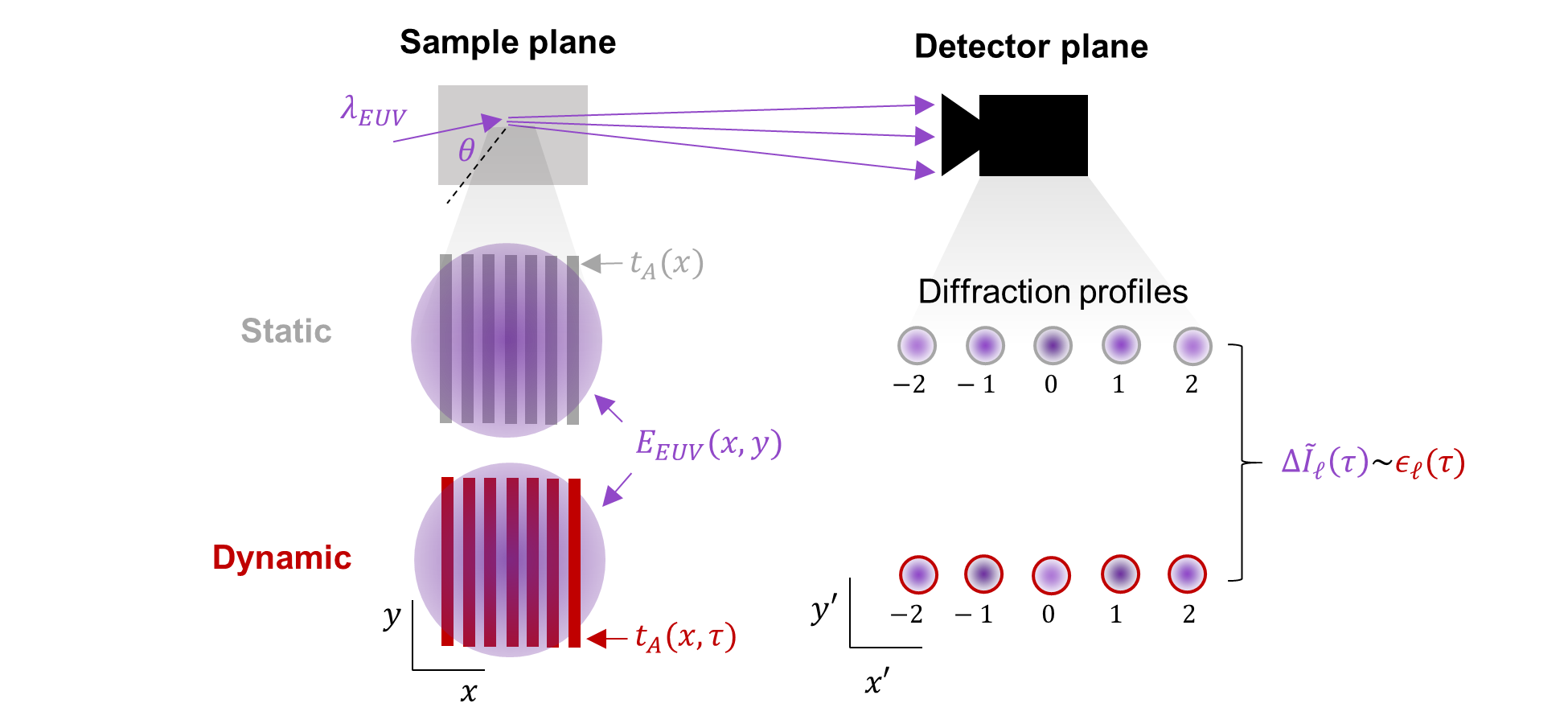}
    \caption{\textbf{Dynamic imaging of periodic structures}. In the sample plane, light diffracts from a periodic profile subject to a small time-dependent perturbation, creating a dynamic diffraction pattern on the detector. Under certain conditions, the change in each far field diffracted order, $\Delta \tilde{I}_{\ell}(\tau)$, can be related to the corresponding Fourier component of the perturbation, $\epsilon_{\ell}(\tau)$.}
    \label{fig:one}
\end{figure}

We begin by considering the diffraction of EUV light from a periodic system subject to a small dynamic perturbation, as shown in Fig. \ref{fig:one}. The physical dynamics may involve thermal, electronic or magnetic processes, resulting in small, measurable variations in the far field EUV diffraction pattern. Using the far field diffraction, we desire to reconstruct the average electric field across a unit cell in the sample plane, in order to spatially resolve the underlying dynamics. Working within the framework of scalar diffraction theory, the electric field of light reflected from the sample is a scalar quantity which shares the period $P$ of the underlying sample and can be expressed at time $\tau$ as 

\begin{equation}
\label{eq:fields}
    E(x,y,z=0,\tau) = E_\text{EUV}(x,y) \cdot t_A(x) \cdot \left[ 1 + \Delta t_A(x,\tau) \right],
\end{equation}
where $x$ and $y$ are the spatial coordinates of the object plane, located at $z=0$. In Eq.~\ref{eq:fields}, $E_\text{EUV}(x,y)$ represents the illumination profile, and $t_A(x)$ and $\Delta t_A(x,t)$ are dimensionless transmittance functions arising from the static periodic profile and small dynamic perturbation, respectively. It is assumed that the static profile has been previously characterized, providing some knowledge of $t_A(x)$, while the unknown $\Delta t_A(x,t)$ is to be reconstructed by measuring small changes in the far field diffraction pattern. Since for small perturbations, $1 + \Delta t_A(x,t) \approx e^{\Delta t_A(x,t)}$, a real or imaginary $\Delta t_A(x,t)$ modifies the electric field amplitude and phase, respectively. For simplicity, it is assumed that the transmittance functions extend infinitely in the ($x$,$y$) plane, such that the finite extent of $E_\text{EUV}(x,y)$ defines the aperture for the diffraction. In the Fresnel approximation within scalar diffraction theory \cite{Goodman2005}, the electric field in the far field, after propagation through a distance $z$, is given by

\begin{equation}
E(x',y',z,\tau) = \frac{e^{ikz}}{i\lambda z}e^{i \frac{k}{2 z} (x'^2 + y'^2)} \int_{-\infty}^{\infty} dx dy e^{-i 2 \pi (f_x x + f_y y)} \left\{E(x,y,\tau)e^{i \frac{k}{2z} (x^2+y^2)}\right\},
\label{eq:fresnel}
\end{equation}
where $\lambda$ and $k = 2 \pi / \lambda$ are the illumination wavelength and wavevector, respectively, $x'$ and $y'$ are the spatial coordinates on the detection plane, and $f_x = x' / \lambda z$ and $f_y = y' / \lambda z$ are the corresponding spatial frequencies. The integral in Eq. \ref{eq:fresnel} is the Fourier transform $\mathcal{F}_{x\rightarrow f_x,y\rightarrow f_y}\left\{\right\}$ of the argument in curly brackets, which consists of the product of four functions and can thus be rewritten using the convolution theorem as 

\begin{equation}
E(x',y',z,\tau) = \frac{e^{ikz}}{i\lambda z}e^{i \frac{k}{2 z} (x'^2 + y'^2)} \left[ \mathcal{F}\left\{e^{i \frac{k}{2z} (x^2+y^2)}\right\} \ast \mathcal{F}\left\{E_\text{EUV}(x,y)\right\} \ast \mathcal{F}\left\{t_A(x)\right\} \ast \mathcal{F}\left\{1+\Delta t_A(x,\tau)\right\} \right],
\label{eq:conv1}
\end{equation}

where the symbol $\ast$ represents a convolution. The third and fourth Fourier transforms in Eq. \ref{eq:conv1} are of the static transmittance function $t_A(x)$ and dynamic perturbation $\Delta t_A(x,\tau)$. These are periodic functions and can therefore be expressed as Fourier series on a discrete set of spatial frequencies, 

\begin{eqnarray}  
\label{eq:fourierseries}
    t_A(x)= \sum_{m = -\infty}^{\infty} \gamma_m e^{i 2 \pi m x/P} \\ \nonumber \Delta t_A(x,\tau) = \sum_{n = -\infty}^{\infty} \epsilon_n(\tau) e^{i 2 \pi n x/P},
 \end{eqnarray} 
where  If $P$ is the spatial period of the sample. In Eq. \ref{eq:fourierseries}, the $\gamma_m$ and $\epsilon_n$ are complex coefficients describing the static profile and dynamic perturbation, respectively. The time-dependent $\epsilon_n(\tau)$ are zero at $\tau=0$, and are assumed to be small for $\tau>0$, such that terms of second order in the $\epsilon_n$ can be neglected. Inserting Eq. \ref{eq:fourierseries} into Eq. \ref{eq:conv1} results in

\begin{eqnarray}\label{eq:conv3}    
    E(x',y',z,\tau) \sim  A(f_x,f_y) \ast \left\{ \sum_{m = -\infty}^{\infty} \gamma_m \delta\left(f_x - \frac{m}{P},f_y \right) \right\} \ast \left\{\delta(f_x,f_y) + \sum_{n = -\infty}^{\infty} \epsilon_n(\tau) \delta\left(f_x - \frac{n}{P} ,f_y \right) \right\},
\end{eqnarray}

where $A(f_x,f_y) = e^{-i\pi\lambda z (f_x^2+f_y^2)}\ast \mathcal{F}\left\{E_\text{EUV}(x,y)\right\}$, and we have omitted the prefactor $e^{ikz}e^{i \frac{k}{2 z} (x'^2 + y'^2)}$, since it does not affect the intensity of the diffracted light. In Eq. \ref{eq:conv3}, $A(f_x,f_y)$ can be convolved with the sums of delta functions inside the curly brackets as $A(f_x-m/P,f_y) \ast \delta(f_x-n/P,f_y) = A(f_x-(n+m)/P,f_y)$, resulting in

\begin{eqnarray}\label{eq:conv4}    
    E(x',y',z,\tau) \sim    \sum_{\ell = -\infty}^{\infty} \gamma_{\ell} A \left(f_x - \frac{\ell}{P} ,f_y\right) + \sum_{m = -\infty}^{\infty} \sum_{n = -\infty}^{\infty} \gamma_m \epsilon_n(\tau) A\left(f_x - \frac{n}{P} - \frac{m}{P},f_y \right).
\end{eqnarray} 
 Eq. \ref{eq:conv4} multiplied by its complex conjugate corresponds to the diffracted intensity measured by a scatterometry experiment in the far field. While this at first appears to be an intimidating operation, many of the resulting cross-terms can be neglected or greatly simplified. In particular, we assume that $E_\text{EUV}(x,y)$ is a sufficiently broad Gaussian function in the sample plane, such that $A(f_x,f_y)$ will contain a sufficiently narrow Gaussian envelope that the diffracted orders are well separated at the detector plane. Under this assumption, $A(f_x-\frac{\ell}{P},f_y) \times A(f_x-\frac{n}{P} - \frac{m}{P},f_y)$ is only nonzero for $\ell = m + n$. Furthermore, any terms of second order in the $\epsilon_n$ can be neglected due to the assumption that the perturbation is small (see Fig.~\ref{fig:four}(a)). The resulting intensity $I = |E|^2$ is

\begin{eqnarray}\label{eq:image}    
    I(x',y',\tau) \approx \sum_{\ell = -\infty}^{\infty} |\gamma_{\ell}|^2 \left|A \left(f_x - \frac{\ell}{P},f_y \right)\right|^2 + \sum_{\ell = -\infty}^{\infty} \sum_{n = -\infty}^{\infty}(\gamma^{*}_{\ell} \gamma_{\ell-n} \epsilon_n(\tau) + \gamma_{\ell} \gamma^{*}_{\ell-n} \epsilon^{*}_n(\tau)) \left|A\left(f_x - \frac{\ell}{P},f_y \right)\right|^2  ,
\end{eqnarray} 
where the $z$ coordinate is omitted from expressions of the intensity at the detector plane for simplicity. The first sum in Eq.~\ref{eq:image} represents the static diffraction profile while the second sum represents the change in diffracted intensity due to the dynamic perturbation. The $\ell$th term in both sums corresponds to the $\ell$th diffracted order, whose intensity is given by

\begin{eqnarray}\label{eq:delimage}    
    I_{\ell}(x',y',\tau) \approx |\gamma_{\ell}|^2 \left|A \left(f_x - \frac{\ell}{P},f_y \right)\right|^2 + \sum_{n = -\infty}^{\infty} (\gamma^{*}_{\ell} \gamma_{\ell-n} \epsilon_n(\tau) + \gamma_{\ell} \gamma^{*}_{\ell-n} \epsilon^{*}_n(\tau)) \left|A\left(f_x - \frac{\ell}{P},f_y \right)\right|^2  ,
\end{eqnarray} 
In Eq. \ref{eq:delimage}, the change in intensity within a particular diffracted order, $\Delta I_{\ell}(x',y',\tau) = I_{\ell}(x',y',\tau) - I_{\ell}(x',y',\tau = 0)$ can be spatially integrated and normalized to the static diffraction profile, 

\begin{eqnarray}\label{eq:delorder}    
    \Delta \tilde{I}_{\ell}(\tau) = \frac{\int dx' dy' \Delta I_{\ell}(x',y',\tau)}{\int dx' dy' I_{\ell}(x',y',\tau=0)},
\end{eqnarray} 
where $\Delta \tilde{I}_{\ell}(\tau)$ is the normalized change in diffraction intensity in order $\ell$. Because the Fresnel diffraction integral in Eq.~\ref{eq:fresnel} makes the paraxial approximation, which assumes that the diffracted rays propagate at small angles relative to the optical axis, it might appear that corrections should be applied to the experimentally measured $\Delta \tilde{I}_{\ell}(\tau)$, especially for near wavelength-scale geometries, before proceeding with the reconstruction. However, as shown in Fig.~1 of the Supplemental Material (SM) \cite{SM}, the normalization in Eq.~\ref{eq:delorder} provides this correction, such that the $\Delta \tilde{I}_{\ell}(\tau)$ are accurate even in non-paraxial geometries. The $\Delta \tilde{I}_{\ell}(\tau)$ and $\epsilon_n$ can be related via a linear system of equations

\begin{eqnarray}\label{eq:linear}    
    \Delta \tilde{I}_{\ell}(\tau) \cdot |\gamma_{\ell}|^2 = \sum_{n=-\infty}^{\infty} [\gamma^{*}_{\ell} \gamma_{\ell-n} \epsilon_n(\tau) + \gamma_{\ell} \gamma^{*}_{\ell-n} \epsilon^{*}_n(\tau)],
\end{eqnarray} 
where the normalization step has eliminated the dependence on $A(f_x,f_y)$, making it unnecessary to precisely characterize the illumination profile when performing a reconstruction. The system of equations can be inverted and solved for the $\epsilon_n$, which are the unknown Fourier components of the perturbation. This is very difficult to accomplish in general for two reasons. First, the matrix to be inverted is not square, and has a left inverse only if its columns are linearly independent. Since each Fourier component has two degrees of freedom corresponding to $\epsilon_n$ and $\epsilon^{\star}_n$, this requires, at a minimum, the measurement of two diffracted orders for each relevant Fourier component of the perturbation.  Second, the matrix which must be inverted in Eq. \ref{eq:linear} consists of the Fourier coefficients ($\gamma_{\ell}$) of the static transmittance function. To invert it, it would be necessary to obtain these coefficients to very high order, which would require an impractically good knowledge of the static sample.

To simplify Eq. \ref{eq:linear} we make the following two assumptions: first that the $\epsilon_n$ are fully imaginary, such that $\epsilon_n = -\epsilon_n^{*}$, and second that

\begin{eqnarray}\label{eq:diagonal}    
    \gamma^{*}_{\ell} \gamma_{\ell-n} = \gamma_{\ell} \gamma^{*}_{\ell-n} \qquad n \neq \ell.
\end{eqnarray} 
In this case, the matrix in Eq.~\ref{eq:linear} is diagonal, such that each Fourier component of the perturbation influences only a single diffracted order as

\begin{equation}\label{eq:linearphasediagonal}    
    \Delta \tilde{I}_{\ell}(\tau) \cdot |\gamma_{\ell}|^2 = [\gamma^{*}_{\ell} \gamma_{0} - \gamma_{\ell} \gamma^{*}_{0} ] \epsilon_{\ell}(\tau).
\end{equation}
Physically, Eq. \ref{eq:linearphasediagonal} applies to an even ($\epsilon_n = \epsilon_{-n}$) phase perturbation or an odd ($\epsilon_n =-\epsilon_{-n}$) amplitude perturbation with respect to a unit cell of the static profile. Because the $\epsilon_n$ are imaginary, an even perturbation implies that $\Delta t_A(x,\tau) = \epsilon_0 + 2 \sum_{n = 1}^{\infty} \epsilon_n(\tau) \cos (2 \pi n x/P)$ affects only phase, while an odd perturbation implies that $\Delta t_A(x,\tau) = 2 i \sum_{n = 1}^{\infty} \epsilon_n(\tau) \sin (2 \pi n x/P)$ affects only amplitude. An even phase perturbation is much easier to realize in experiments and is a useful approximation in cases where the static geometry is also even, the excitation is uniform, and the perturbation consists mainly of phase changes caused by surface deformation as opposed to a change in reflectivity.

The second condition given by Eq. \ref{eq:diagonal} is satisfied when for any two indices, $\ell, \ell' \neq 0$, the product $\gamma^{*}_{\ell} \gamma_{\ell'}$ is real. This in turn requires that all $\gamma_{\ell \neq 0}$ have the same phase, mod($\pi$), or equivalently that $\gamma_{\ell'} = a\gamma_{\ell}$, where $a$ is a real number. It is important that $\gamma_0$ does not also meet this condition, otherwise the entire matrix will be zero and the perturbation will have no effect on the diffraction efficiencies to first order. While this condition may seem obscure, it is shown in section \ref{sec:three} that it applies to the case of a typical square grating.

The form of Eq.~\ref{eq:linearphasediagonal} is extremely convenient because the reconstruction of a particular Fourier component $\epsilon_{\ell}(\tau)$ of the perturbation only requires knowledge of the corresponding $\ell$th component of the static profile and the change in the $\ell$th diffracted order. However, Eq.~\ref{eq:linearphasediagonal} is only valid for a small even phase or odd amplitude perturbation to a well-behaved static transmittance function in the framework of scalar diffraction theory. The following three sections examine the consequences of these assumptions and the performance of the reconstruction in scenarios where they are not fully met. 

\section{Numerical simulations for a thin square grating}
\label{sec:three}

The following section presents numerical simulations of an idealized thin square grating where all of the assumptions leading to Eq.~\ref{eq:linearphasediagonal} hold, permitting an accurate reconstruction of the unit cell dynamics. In scalar diffraction theory, the transmittance function of a thin square grating can be written as 

\begin{align}
t_A(x) = 
\begin{cases}
        r_g e^{-i\phi_H} & |x|\leq\frac{L}{2} \\
        r_s & \frac{L}{2} <|x| < \frac{P}{2}, 
\label{eq:fieldgtg}
\end{cases}
\end{align}
\begin{figure}
    \includegraphics[width=1.0 \columnwidth]{Fig2_v3.png}
    \caption{\textbf{Reconstruction of a dynamic phase perturbation to a square grating} (a) The static grating geometry consists of a 1D array of rectangular structures of linewidth $L$, height $H$ and period $P$ on a substrate (see Eq. \ref{eq:fieldgtg}). The complex EUV reflectivities of the grating and substrate are $r_g$ and $r_s$, respectively. The transmittance function for the static grating (blue) is multiplied by a small dynamic phase perturbation (red), which is exaggerated here. (b) The Fourier components $\epsilon_{\ell}$ of the perturbation (red) and their reconstructions (green) from a Fresnel simulation of the far field diffraction pattern. Each Fourier component of the perturbation is reconstructed using the corresponding far field diffracted order. (c) By including an increasingly large number of diffracted orders in the reconstruction (labeled numerically) and Fourier transforming back into real space, one approaches the original dynamic perturbation.}
    \label{fig:two}
\end{figure}
where $P$ is the grating period, $L$ is the grating linewidth, and $r_g = R_g e^{i \phi_g}$ and $r_s = R_s e^{i \phi_s}$ are the grating and substrate reflectivities. The grating reflectivity takes into account the multilayer stack of the grating and underlying substrate material. The geometric phase $\phi_H = 4\pi H \cos{\theta} / \lambda$ arises from the change in optical path length due to the grating height $H$ for light of wavelength $\lambda$ and incidence angle $\theta$ from normal. The reflectivities $r_s$ and $r_g$ are assumed to be constants, such that any dynamic perturbation consists of a surface displacement which only affects the phase of the transmittance function. Because the static unit cell is symmetric about $x$ = 0, we expect that any spatially uniform excitation will produce an even perturbation. The static transmittance function given by Eq. \ref{eq:fieldgtg} can be written as a complex Fourier series, with coefficients 

 \begin{eqnarray}  
\label{eq:gtgfouriercomponents}
    \gamma_{m \neq 0} = \frac{L}{P} (r_g e^{-i \phi_H} - r_s) \text{sinc}\left(\frac{mL}{P} \right) \\ \nonumber 
    \gamma_0 = \frac{L}{P}(r_g e^{-i \phi_H}) + r_s \left( 1- \frac{L}{P} \right),
 \end{eqnarray} 
 where $\text{sinc}(x) = \sin(\pi x) / (\pi x)$. Because all of the $\gamma_{m \neq 0}$ have the same phase mod($\pi$), Eq. \ref{eq:diagonal} is satisfied and the reconstruction can proceed using Eq. \ref{eq:linearphasediagonal}. Inserting Eq. \ref{eq:gtgfouriercomponents} into Eq. \ref{eq:linearphasediagonal} yields the relation
 
\begin{equation} \label{eq:gtgreconstruction}
	\Delta \tilde{I}_{\ell}(\tau) = i \epsilon_{\ell}(\tau) \frac{2P}{L} \left[ \text{sinc}\left( \frac{\ell L}{P}\right) \right ]^{-1} \frac{R_gR_s\sin(\phi_s+\phi_H-\phi_g)}{R_g^2 + R_s^2 - 2 R_g R_s \cos(\phi_s+\phi_H-\phi_g)} ,
\end{equation}
which enables the reconstruction of the Fourier components $\epsilon_{\ell}(\tau)$ from the normalized changes in diffraction intensity $\Delta \tilde{I}_{\ell}(\tau)$. Fig. \ref{fig:two}(a) shows an example unit cell of a square-wave grating, and a simulated static transmittance function (blue) and phase perturbation (red). In Fig.~\ref{fig:two}(b), the Fourier components of the phase perturbation (red) are compared to their reconstructed values (green) obtained using Eq. \ref{eq:gtgreconstruction} from a Fresnel diffraction simulation of the far field diffraction pattern. The Fresnel diffraction simulation assumes a Gaussian beam of 25~\textmu m radius and 30~nm wavelength incident normally on a 1000~nm period grating,  and a 10~cm sample to camera distance. The very good agreement shown is contingent on the phase perturbation being small, \textit{i.e.}, $\Delta t_A(x,\tau) \ll 1$, such that terms of second order in the $\epsilon_{\ell}$ may be neglected. Fig.~\ref{fig:two}(c) shows the reconstruction in real space as compared to the original perturbation. The numbers below the curves indicate the number of Fourier components, each corresponding to one diffracted order, used in the reconstructions. With the increasing spatial frequency content provided by including additional diffracted orders, the reconstruction improves and approaches the original perturbation.

\section{Dynamic amplitude perturbations}
\label{sec:four}

In many situations, phase and amplitude perturbations appear simultaneously, making it necessary to consider both when attempting to apply the methods above. For example, in an EUV scatterometry experiment where the wavelength is chosen to be off-resonance with electronic processes, the experimental signal is sensitive only to surface displacements and material density changes caused by changes in phonon temperature. While a vertical surface displacement will produce a phase perturbation, a change in density will generally result in both an amplitude and phase perturbation. As discussed in section \ref{sec:two}, an amplitude perturbation can only be simplified in the same way as a phase perturbation if it is odd with respect to the unit cell and unaccompanied by a phase change. Since simultaneous amplitude and phase perturbations in experiments will generally be spatially correlated, one must instead return to the matrix inversion in Eq. \ref{eq:linear}, which is in general very difficult to solve.

\begin{figure}
    \centering
    \includegraphics[width=1.0\linewidth]{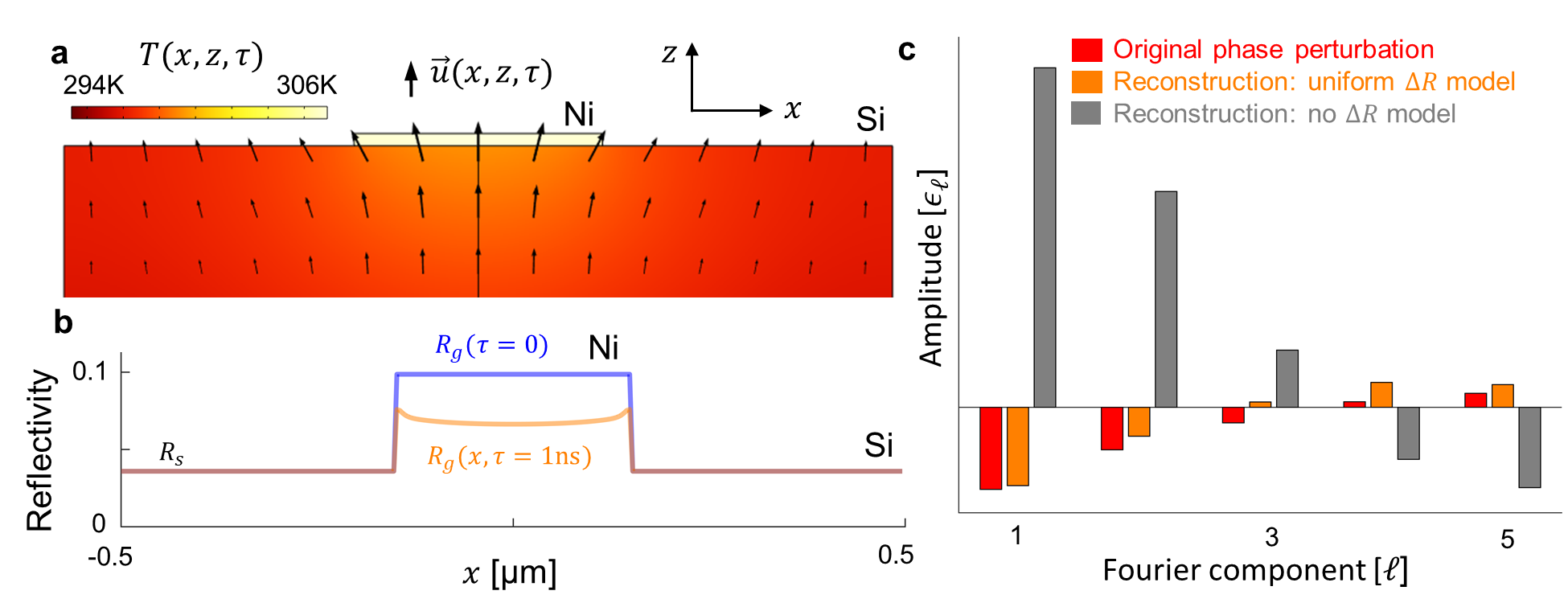}
    \caption{\textbf{Simultaneous treatment of an amplitude and phase perturbation in the square grating geometry} (a) Finite element calculation of the temperature $T(x,z,\tau)$ and displacement $\vec{u}(x,z,\tau)$ fields in a unit cell of the square nickel grating geometry one nanosecond after the nickel structure is excited by an ultrafast pulse. (b) EUV reflectivity across one unit cell before (blue) and after (orange) the excitation. The change in reflectivity is exaggerated for visibility. While the change in silicon substrate reflectivity is assumed to be negligible due to its low temperature rise, the change in nickel reflectivity varies laterally as a function of density. (c) Fourier components $\epsilon_{\ell}$ of the phase perturbation calculated directly from the simulated vertical displacement using Eq. \ref{eq:phasetoheight} (red) and their reconstruction from a Fresnel simulation of the far field diffraction pattern under two scenarios. The reconstruction performs reasonably well when using  Eq. \ref{eq:gtgreconstructionamp} to account for the change in reflectivity (orange). If the reflectivity is assumed to be static, as in Eq. \ref{eq:gtgreconstruction}, the reconstruction fails (gray).}
    \label{fig:three}
\end{figure}

Fortunately, in some contexts it is possible to use the overall change in reflectivity combined with knowledge of the system under study to separately analyze the amplitude and phase perturbations and still achieve an approximate spatial reconstruction of the latter. As an example, we consider a finite element simulation of an ultrafast laser pulse of $\sim$800~nm wavelength incident on the grating geometry in Fig.~\ref{fig:two}(a). The grating linewidth ($L$), period ($P$) and height ($H$) are set to 300~nm, 1000~nm and 15~nm, respectively and the grating substrate materials are chosen to be nickel and silicon, respectively. For near infrared light, the absorption mainly occurs in the nickel, where excited electrons equilibrate with the phonons on a roughly picosecond timescale \cite{Longa2007}. Since we are interested in reconstructing the dynamic surface deformation at much later times, it is only necessary to model the evolution of a single lattice temperature after the injection of a spatially uniform, picosecond-scale heat pulse into the nickel. We assume a typical thermal boundary resistance between the nickel and silicon of 5  nK$\cdot$m$^2$/W \cite{Cheaito2015}. The finite element simulation, implemented in COMSOL \cite{comsol}, calculates the dynamic temperature $T(x,z,\tau)$ and displacement $\vec{u}(x,z,\tau)$ fields by solving the thermal diffusion equation and elastic equations of motion with a linear thermoelastic coupling, within a unit cell geometry subject to periodic boundary conditions \cite{JoshThesis}. The calculation is performed in the quasi-static approximation, where $\partial^2\vec{u} / \partial \tau^2$ is set to zero in the equation of motion, such that the displacement is solely the result of thermal expansion rather than acoustic oscillations launched by the short pulse. Table \ref{tab:one} shows the parameters used in the simulation and Fig. \ref{fig:three}(a) shows the temperature profile and displacement one nanosecond after the excitation. Because the nickel absorbs far more of the infrared pump pulse per unit volume than the silicon and because of the high thermal boundary resistance between the two materials, the change in nickel temperature is always far higher than the change in silicon temperature. It is therefore reasonable to neglect any change in silicon density and treat the substrate reflectivity as a constant, making it only necessary to account for the change in nickel density and the surface displacement in the reconstruction. 

\begin{table}[hbt!]
\centering
\begin{tabular}{ |p{3.6cm}|p{2cm}|p{2cm}|p{2cm}|p{2cm}| }
 \hline
 \textbf{Material} & $\rho$ [g/cm$^3$] & $\kappa$ [W/m/K] & $\text{c}_\text{p}$ [J/kg/K] & $\alpha$ [1/K] \\
 \hline
 Grating (Ni) & 8.9 & 90.9 & 444 & 12.8e-6 \cite{Kollie1977}  \\
 \hline
 Substrate 1 (Si) & 2.33 & 149 & 712 & 2.6e-6 \cite{Okada1984} \\
 \hline
  Substrate 2 (Diamond) & 3.52 & 2300 & 508 & 1e-6 \cite{Stoupin2011} \\
 \hline
\end{tabular}
\caption{\textbf{Finite element model material parameters.} The critical material parameters are density ($\rho$), specific heat capacity ($\text{c}_\text{p}$), thermal conductivity ($\kappa$) and linear coefficient of thermal expansion ($\alpha$). The values in the first three columns come from Ref. \cite{Lide2005}. The elastic constants also appear in the model \cite{Hopcroft2010,Klein1993,Nardi2011,Hoogeboom-Pot2016} but have little impact in the quasi-static approximation. The values for diamond are used in simulations of the experiment in section~\ref{sec:experiment}.}
\label{tab:one}
\end{table}

To proceed with the reconstruction, it is necessary to calculate the complex static reflectivities $r_s$ and $r_g$ of the substrate and the grating on substrate stack, respectively. The EUV refractive indices of each material can be calculated using wavelength-dependent elemental scattering factors tabulated by the Center for X-ray Optics and material densities \cite{Henke1993,Attwood2016}. The reflectivities are then calculated using the Parratt formalism \cite{Parratt1954}. In scalar diffraction theory, the static transmittance function at the object plane is a function of $r_s$, $r_g$ and the grating height, as given by Eq. \ref{eq:fieldgtg}. The phase perturbation due to the vertical component of the displacement, $u_z(x,\tau)$, is given by

\begin{equation}\label{eq:phasetoheight}
    \Delta t_A(x,\tau) = -i \frac{4 \pi \cos\theta}{\lambda} u_z(x,\tau),
\end{equation}

where $\theta$ is the EUV incidence angle relative to normal. To further simplify this example calculation, we neglect the horizontal expansion of the structure. In the absence of an amplitude perturbation or significant horizontal displacement, Eqs. \ref{eq:gtgreconstruction} and \ref{eq:phasetoheight} allow for a direct reconstruction of the vertical displacement. However, the finite element simulation and subsequent reflectivity calculation reveal that the nickel density and height change sufficiently that $r_g(x,\tau)$  varies both temporally and spatially across the grating surface. In general, it is difficult to simultaneously reconstruct $r_g(x,\tau)$ and $u_z(x,\tau)$; however, the application of certain physical constraints based on the geometry and heating profile can make the problem much more tractable. For example, one might assume that $r_g(x,\tau)$ and $u_z(x,\tau)$ are proportional at each location $x$ and perform an iterative reconstruction. Alternatively, we can assume that any spatial variation in $r_g(x,\tau)$ across the grating surface is small compared to the spatially-averaged change relative to the static value,

\begin{equation}\label{eq:gtgreflectivity}
    r_g(x,\tau) \approx r_g(\tau) = [R_g + \Delta R_g(\tau)]e^{i[\phi_g + \Delta \phi_g(\tau)]}.
\end{equation}
In Eq. \ref{eq:gtgreflectivity}, $\Delta R_g(\tau)$ and $\Delta \phi_g(\tau)$ are real numbers, representing the spatially-averaged change in the amplitude and phase of the complex reflectivity, respectively, from their static values $R_g$ and $\phi_g$. The static diffraction intensity from a square grating can be expressed in terms of $R_g$ and $\phi_g$ using Eq.~\ref{eq:image} and Eq.~\ref{eq:gtgfouriercomponents} and then expanded to first order in $\Delta R_g(\tau)$ and $\Delta \phi_g(\tau)$. Normalizing this result and combining with Eq. \ref{eq:gtgreconstruction} yields

\begin{align}  
\label{eq:gtgreconstructionamp}
\Delta  \tilde{I}_{\ell}(\tau) = 2 &[R_g^2 + R_s^2 - 2 R_g R_s \cos(\Phi)]^{-1} \times \\ \nonumber &  \left\{ i \epsilon_{\ell}(\tau) \frac{P}{L} \left[ \text{sinc}\left( \frac{\ell L}{P}\right) \right ]^{-1} R_gR_s\sin(\Phi) + \Delta R_g(\tau)(R_g - R_s \cos(\Phi))-\Delta \phi_g(\tau)(R_g R_s \sin(\Phi))  \right\} ,
 \end{align} 

where $\Phi = \phi_s+\phi_H-\phi_g$ and terms of second order in the small quantities $\epsilon_{\ell}(\tau)$, $\Delta R_g(\tau)$ and $\Delta \phi_g(\tau)$ have been neglected. In Eq. \ref{eq:gtgreconstructionamp}, $\Delta R_g(\tau)$ and $\Delta \phi_g(\tau)$ do not vary between diffracted orders. They can be accounted for by considering the total normalized change in reflectivity, defined as $\Delta \tilde{I}(\tau) = \int dx' dy' \Delta I(x',y',\tau) / \int dx'dy' I(x',y',\tau=0)$ in analogy to Eq. \ref{eq:delorder}. For the case of a square grating, $\Delta \tilde{I}(\tau)$ depends only on $\Delta R_g(\tau)$ and static parameters,

\begin{equation} \label{eq:delrg}
	 \Delta \tilde{I}(\tau) = \Delta R_g(\tau) \frac{2 R_g}{\left(\frac{P}{L}-1\right)R_s^2 + R_g^2}.
\end{equation}

Eq. \ref{eq:delrg} can be derived by observing that $\int dx'dy' I(x',y',\tau=0) \sim \frac{L}{P}R_g^2 + (1-\frac{L}{P})R_s^2$ is independent of $\phi_g$ and expanding it to first order in $\Delta R_g(\tau)$. Experimentally, determining $\Delta R_g(\tau)$ in this way requires a large enough numerical aperture such that any diffracted orders not captured on the camera contribute only negligibly. The remaining variable, $\Delta \phi_g(\tau)$, can be calculated from $\Delta R_g(\tau)$ by parameterizing both as functions of atomic density.

The approximation of $r_g(x,\tau)$ as spatially uniform still permits a fairly accurate reconstruction of the phase perturbation in numerical simulations. This is possible in part because, as shown in Fig. \ref{fig:three}(b), the spatial variation in $r_g(x,\tau)$ calculated from the finite element results is relatively minor in comparison to the average change relative to the unperturbed reflectivity. Fig. \ref{fig:three}(c) compares the $\epsilon_{\ell}$ components of the phase perturbation obtained using Eq. \ref{eq:phasetoheight} from the vertical displacements calculated by the finite element simulation one nanosecond after the excitation (red), to their reconstructions from a simulated far field diffraction profile under two scenarios. In the first scenario (orange), the Fresnel diffraction accounts for both the vertical surface displacement (phase perturbation) and the spatial variation in $r_g(x,\tau)$, due to the change in nickel thickness and density as a function of $x$, integrated across $z$. The parameter $\Delta R_g(\tau)$ is then calculated from the simulated diffraction profile using Eq.~\ref{eq:delrg} and the corresponding value of $\Delta \phi_g(\tau)$ is obtained from $\Delta R_g(\tau)$ by relating both to an average change in nickel density. The reconstruction then proceeds via Eq.~\ref{eq:gtgreconstructionamp}. The reconstruction is not as accurate as the results in Fig.~\ref{fig:two}(b), because the spatially-varying grating reflectivity appears in the Fresnel diffraction simulation, while the reconstruction assumes it is uniform. The second scenario (gray) uses the same Fresnel diffraction simulation as the first scenario, but performs the reconstruction using Eq.~\ref{eq:gtgreconstruction}, thereby neglecting the amplitude perturbation altogether. The failure of the reconstruction in the second scenario serves as a control to illustrate the necessity of accounting for an amplitude perturbation if it is present. The example shown here illustrates how physical knowledge of a particular system and its symmetries can allow for the handling of more complex dynamics in the reconstruction.

\section{Robustness of the reconstruction method}
\label{sec:five}

\begin{figure}
    \centering
    \includegraphics[width=0.9\linewidth]{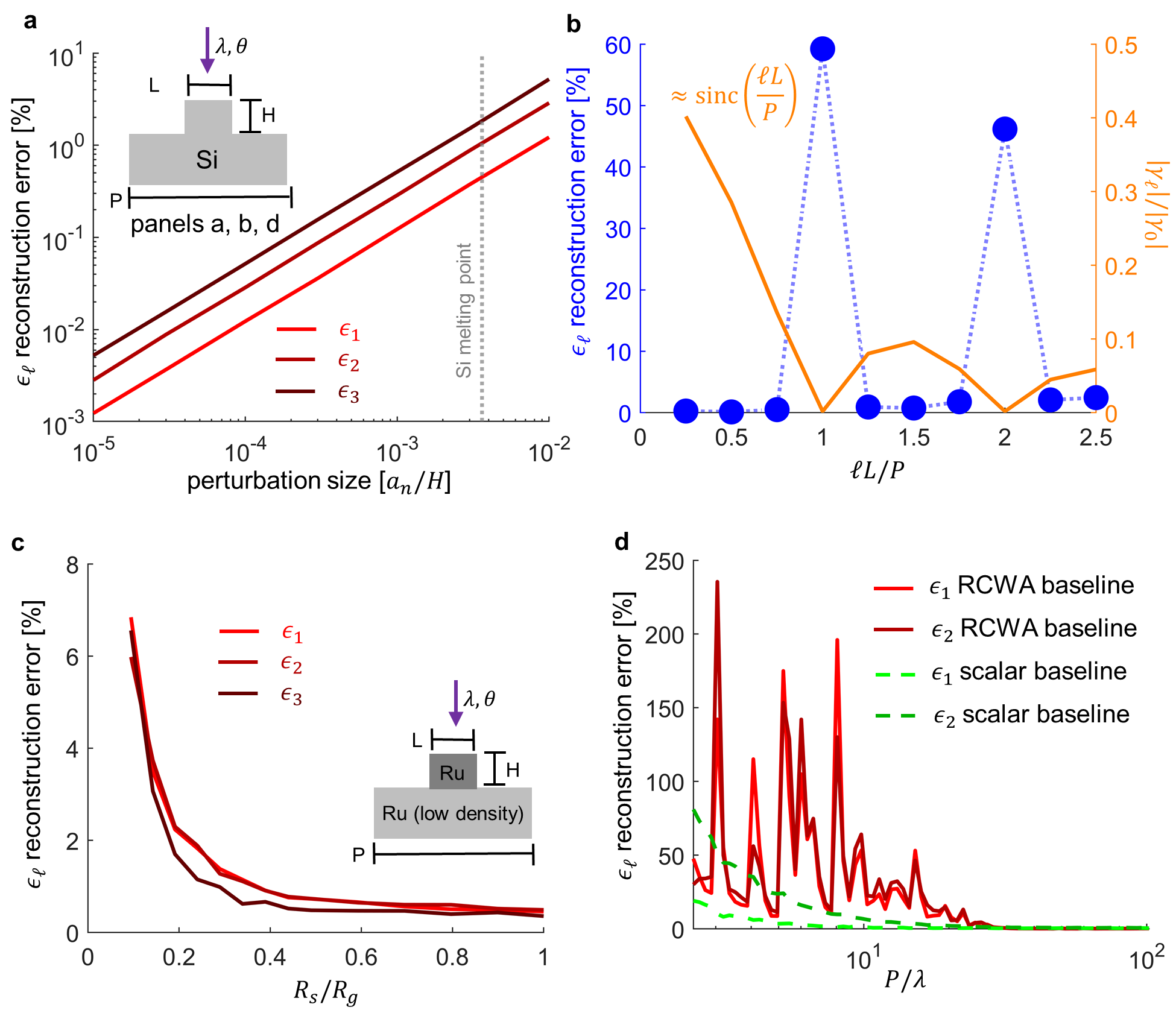}
    \caption{\textbf{Simulated reconstruction method robustness} (a) Average error in the reconstructed Fourier coefficients $\epsilon_1$ to $\epsilon_3$ of the transmittance function perturbation $\Delta t_A(x,\tau)$ as a function of perturbation size. The simulation considers $\lambda$ =  30~nm EUV light normally incident on a square silicon grating of $H = 3\lambda/8$, $L$ = 500~nm and $P$ = 2000~nm (inset) and computes random surface deformations of the form in Eq.~\ref{eq:randompert} with Fourier coefficients chosen from a uniform distribution over [-$a_n$,$a_n$]. The corresponding transmittance function perturbations are directly computed using Eq.~\ref{eq:phasetoheight} and compared to the results of a Fraunhofer diffraction simulation and subsequent reconstruction. The error grows as a function of perturbation size because Eq.~\ref{eq:linearphasediagonal} neglects terms of second order in the $\epsilon_{\ell}$. The error here and in subsequent panels is defined as the percentage difference between the reconstructed and directly computed Fourier coefficients of the transmittance function perturbation averaged over many random trials. (b) Average reconstruction error vs. Fourier coefficient for a small perturbation to the geometry in panel (a). The random surface deformations are computed as before, but now the object plane transmittance function and far field diffraction pattern are computed using an RCWA simulation, which introduces small off-diagonal terms neglected in scalar diffraction theory. When the static diffraction efficiency is also small, which for a square grating is the case when $\ell L/P$ is an integer, the reconstruction error is large. (c) Average error in reconstructed Fourier coefficients as a function of $R_s/R_g$. To isolate the effects of reflectivity, the simulation considers 25~nm EUV light normally incident on a square ruthenium (Ru) grating of $H = 5 \lambda / 8$, $L$ = 200~nm and $P$ = 2000~nm and twice the nominal density, so that the structure transmits no light, on a Ru substrate of variable density and reflectivity. The same RCWA simulation is used as in panel (b) and the error grows as $R_s/R_g$ decreases, because this causes the size of the on-diagonal term in Eq.~\ref{eq:linear} to decrease. (d) Average error in the reconstructed Fourier coefficients $\epsilon_1$ and $\epsilon_2$ as a function of grating size using the method of generating random surface deformations described in panel (a). The simulation considers $\lambda$ = 30~nm EUV light incident normally on a silicon grating of $H = 3\lambda/8$ and variable period $P$, with $L = 0.3P$. As in panels (b) and (c), the reconstruction is performed on an RCWA simulation of the far field diffraction. For the solid red curves, the  $\gamma_{\ell}$ used in the reconstruction and the perturbed transmittance function to which to compare the reconstruction results are calculated directly using RCWA. For the dashed green curves the $\gamma_{\ell}$ are computed using Eq.~\ref{eq:gtgfouriercomponents} and the reconstruction is compared to the perturbed transmittance function calculated from the vertical surface displacements using Eq.~\ref{eq:phasetoheight}.}
    \label{fig:four}
\end{figure}

The framework presented thus far works for small perturbations and relies on typical approximations in Fresnel diffraction theory, in particular the Kirchhoff approximation within the framework of scalar diffraction theory, which neglects the vector nature of the $\vec{E}$ and $\vec{H}$ fields, and represents the scalar field at each point on the sample plane by the value it would assume in the absence of any boundaries. This section examines situations in which these conditions are only approximately met and the implications for the reconstruction method. In particular, as shown in Fig.~\ref{fig:four}, the perturbations need to remain small, and off-diagonal terms in Eq.~\ref{eq:linear} need to remain nearly zero.

If the perturbation to the transmittance function is sufficiently large, terms of second order in $\epsilon_{\ell}$ can no longer be neglected as in Eq. \ref{eq:image}. To determine the magnitude of the error introduced as a function of perturbation size, we consider 30~nm light normally incident on a square silicon grating of $L$=500~nm and $P$=2000~nm as shown in the inset to Fig. \ref{fig:four}(a). In this simple case, the static profile is a pure phase grating of maximum contrast, with $\phi_H = 3 \pi / 2$ such that $H = 3 \lambda / 8$. We next simulate random vertical surface displacements to the static profile of the form

\begin{equation} \label{eq:randompert}
	 u_z(x) = \sum_{n=1}^N \tilde{a}_n \cos(2 \pi x n /P),
\end{equation}
where the $\tilde{a}_n$ are random variables chosen from the uniform distribution [$-a_n$,$a_n$]. To make the perturbation assume a physically realistic shape rather than appear as random noise, the size of each Fourier coefficient in Eq. \ref{eq:randompert} is set to decrease as $a_n$ = $a_1 n^{-1/2}$ up to term $N=15$ (see Fig. 2 in the SM \cite{SM}). The vertical surface displacement changes the phase of the transmittance function, and the Fourier components $\epsilon_{\ell}$ of this phase perturbation can be directly calculated using Eqs. \ref{eq:fourierseries} and \ref{eq:phasetoheight} and compared to the reconstructed values to determine the reconstruction error. Using the static and perturbed transmittance functions, the change in diffraction efficiency of each order in the far field is calculated using a Fraunhofer diffraction simulation for simplicity, assuming an incident plane wave. Finally, the $\epsilon_{\ell}$ are reconstructed using Eq.~\ref{eq:linearphasediagonal}. To define a metric for the average reconstruction error, we first calculate the average absolute difference between the reconstructed and directly calculated $\epsilon_{\ell}$ across many random perturbations and then divide this quantity by the average magnitude of the calculated $\epsilon_{\ell}$. As shown in Fig. \ref{fig:four}(a), the typical reconstruction error grows linearly with perturbation size and trends to zero for small perturbations, regardless of which order of the perturbation is being reconstructed.

The remainder of the present section is dedicated to examining situations in which Eq.~\ref{eq:linear} is only approximately diagonal. If the off-diagonal terms are present, but small, then the reconstruction will be relatively accurate; however, if the off- and on-diagonal terms are of similar size, then it is necessary to solve a full linear system of equations. As discussed above, this presents a major problem because it requires the collection of many diffracted orders and a complete knowledge of the static profile in order to reconstruct each individual Fourier component of the perturbation. The square silicon grating geometry shown in the inset to Fig. \ref{fig:four}(a) again serves as the test system, subject to random vertical displacements as defined in Eq.~\ref{eq:randompert}. A simulation using the RETICOLO implementation of rigorous coupled-wave analysis (RCWA) \cite{reticolo} first directly calculates the Fourier components, $\gamma_{\ell}$ and $\epsilon_{\ell}$, of the static transmittance function and small perturbation due to the surface displacement, respectively. This is accomplished by calculating the reflected electric field some distance above the sample, where evanescent modes are negligible, and then back-propagating it using the angular spectrum method \cite{Goodman2005} to the plane defined by the top of the grating. The RCWA calculation for the perturbed geometry relies on a fine pixelation near the grating surface to accommodate the very small deformation. Next, the RCWA simulation calculates the normalized change in intensity in each diffracted order. The $\epsilon_{\ell}$ are then reconstructed using Eq.~\ref{eq:linearphasediagonal} and the $\gamma_{\ell}$ calculated using RCWA. The same metric is applied as in Fig.~\ref{fig:four}(a) to determine the average error between the reconstructed and calculated $\epsilon_{\ell}$ over a number of random perturbations. It is important to note that for a square grating in scalar diffraction theory, the $\gamma_{\ell}$ given in Eq.~\ref{eq:gtgfouriercomponents} ensure that the off-diagonal terms in Eq.~\ref{eq:linear} are all zero; it is only through non-scalar effects in the RCWA simulations that off-diagonal contributions appear. Fig. \ref{fig:four}(b) (blue curve) shows the reconstruction error as a function of Fourier coefficient ($\epsilon_{\ell}$) for a $L$=500~nm, $P$=2000~nm grating. The error spikes when $\ell L/P$ is an integer because the on-diagonal term is small, such that it is overwhelmed even by small off-diagonal contributions. Physically, this corresponds to orders with a low static diffraction efficiency. In the Fresnel approximation, the static diffraction efficiency at integer $\ell L/P$ is exactly zero, due to the sinc function in Eq.~\ref{eq:gtgfouriercomponents}. The orange curve in Fig. \ref{fig:four}(b), plots the magnitude of $|\gamma_{\ell}|/|\gamma_0|$ calculated by the RCWA simulation, which is nonzero but still very small at integer $\ell L/P$. The reconstruction also suffers even when the static diffraction efficiency is large if $\gamma_0$ has the same phase as the remaining $\gamma_{\ell}$. For example, if $R_g \gg R_s$ for a square grating, then, using Eq.~\ref{eq:gtgfouriercomponents}, $\gamma_0 \approx \frac{L}{P}r_g e^{-i \phi_H}$, $\gamma_{\ell \neq 0} \approx \frac{L}{P} r_g e^{-i \phi_H} \text{sinc}\left(\frac{\ell L}{P}\right)$ and the on-diagonal term $\gamma^{*}_{\ell} \gamma_{0} - \gamma_{\ell} \gamma^{*}_{0}$ is approximately zero. Fig. \ref{fig:four}(c) isolates this effect by considering a square ruthenium (Ru) grating where the substrate density and reflectivity varies relative to that of the structure (see inset). For 25~nm light normally incident on a $H = 5 \lambda / 8$~nm Ru structure of twice the nominal density, virtually no light reaches the substrate, ensuring that only $R_s$ changes when the substrate density is varied. Using the same RCWA calculation and reconstruction of the perturbed transmittance function as above, we observe that the average reconstruction error rises as $R_s/R_g$ decreases. Physically, when all of the Fourier components of the static transmittance function have the same phase, including $\gamma_0$, there is no interference to first order between static profile and dynamic phase perturbation and only terms of second order in the $\epsilon_{\ell}$ can appear.

Scalar diffraction theory and the Kirchhoff approximation assume that the aperture size in the object plane is much longer than a wavelength, which for a grating means $L \gg \lambda$ and $P-L \gg \lambda$, such that the reflectivity at each point on the grating or substrate is independent of its proximity to the structure edges. In reality, the electric field at distances on the order of a wavelength from the edge of a structure will be influenced by the boundary and this will create small ripples in the transmittance function. We therefore expect that the reconstruction will perform poorly for very small gratings, since there will be significant off-diagonal terms. For wavelength-scale structures, it is also no longer accurate to convert between surface displacement and the perturbed transmittance function using Eq.~\ref{eq:phasetoheight}; this raises the question of whether one should attempt to reconstruct the true $\epsilon_{\ell}$ or their values predicted by scalar diffraction theory. Fig.~\ref{fig:four}(d) illustrates these considerations using the same silicon geometry as in panels (a) and (b), with $H = 3\lambda/8$, $L$= 0.3$P$ and $P$ varied relative to $\lambda$=30~nm. The red and green curves show two methods of reconstructing $\epsilon_1$ and $\epsilon_2$ from the same set of random surface displacements and RCWA diffraction simulations. Both approaches encounter difficulties for $P \lesssim 10 \lambda$ but exhibit dramatically different behavior at the smallest length scales. For the red curves, RCWA is used to directly calculate both the $\gamma_{\ell}$ used the reconstruction and the $\epsilon_{\ell}$ for comparison with the reconstruction results. The error spikes occur because at small length scales the vertical displacement affects both the amplitude and phase of the calculated transmittance function, while the reconstruction assumes a pure phase perturbation. The green curves use the analytic expressions for the $\gamma_{\ell}$ given by Eq.~\ref{eq:gtgfouriercomponents} in the reconstruction and compare the reconstructed $\epsilon_{\ell}$ to values calculated directly from the surface displacements using Eq.~\ref{eq:phasetoheight}. The first approach (red curves) may be more appropriate when the goal is to calculate the transmittance function, while the second approach (green curves) may be more useful when the goal is to obtain information about the sample dynamics. The two approaches break down for different reasons for wavelength-scale geometries, illustrating how the method of reconstructing the $\epsilon_{\ell}$ and relating them to the sample dynamics is a very context-dependent problem which becomes more complicated at smaller scales.

\section{Experimental reconstruction of average unit cell dynamics using EUV pump-probe scatterometry}
\label{sec:experiment}

\begin{figure}
    \centering
    \includegraphics[width=0.9\linewidth]{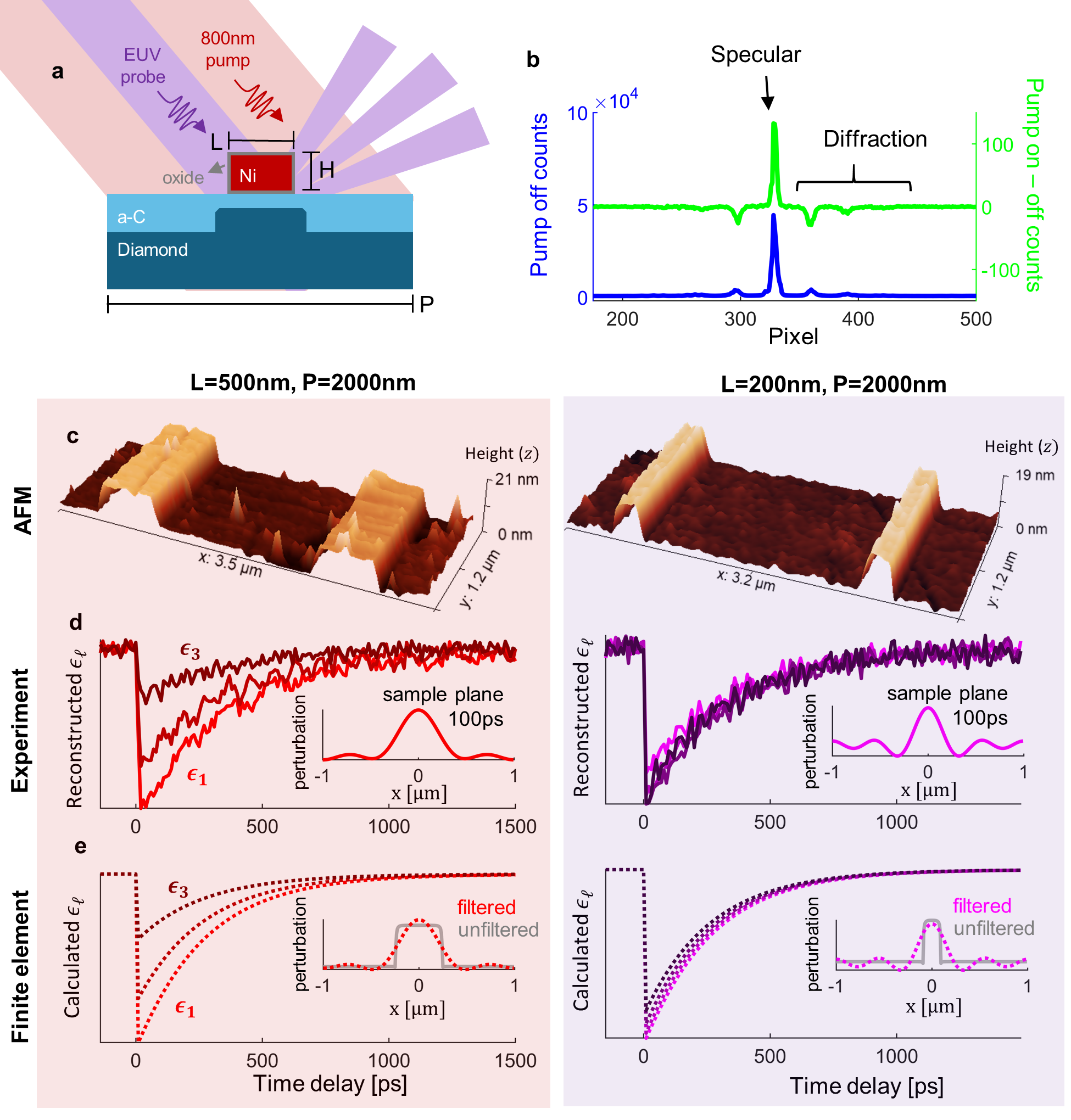}
    \caption{\textbf{Experimental reconstruction of unit cell dynamics for a nickel grating on diamond} (a) Schematic of the dynamic EUV scatterometry experiment and one unit cell of the nickel grating on the diamond substrate with oxide and amorphous carbon interlayer. (b) Vertically binned EUV diffraction pattern on the CCD camera with the pump blocked (blue). Difference in counts with the pump unblocked versus blocked (green) 100~ps after the excitation. (c) Example AFM scans for the nominal $L$=500~nm, $P$=2000~nm and $L$=200~nm, $P$=2000~nm gratings used to extract linewidth, height and surface roughness. (d) Reconstructed $\epsilon_{\ell}$ as a function of pump-probe time delay for both gratings. Each curve shows a different $\epsilon_{\ell}$ from low (bright) to high (dark). The positive and negative diffracted orders are averaged in the reconstruction. The insets show the vertical surface displacement ($u_z$) calculated using Eq. \ref{eq:phasetoheight} at a 100~ps delay. (e) The corresponding finite element calculation of $\epsilon_{\ell}$ using Eq. \ref{eq:phasetoheight}, normalized to the same average value as the experimental reconstruction. The insets show the vertical surface displacement at 100~ps. Gray curves show the full finite element simulation, while red and purple curves show the result of a low pass spatial filter capturing only those Fourier components which can be reconstructed experimentally.}
    \label{fig:five}
\end{figure}

In this final section, we perform an example reconstruction using the dynamic EUV scatterometry experiment discussed in Refs. \cite{Frazer2019,McBennett2024}. Fig. \ref{fig:five}(a) shows the experimental setup, which begins with the output of a 4~kHz, 25~fs Ti:Sapphire amplifier. The 800~nm pulses are split into a pump and probe beam. The pump beam passes through a linear delay stage allowing for its arrival time at the sample plane to be adjusted relative to the probe, while the probe beam undergoes high harmonic generation in an argon-filled waveguide to produce coherent EUV light of approximately 30~nm wavelength, as confirmed by careful measurements of the sample to detector distance and a fit of the static diffraction profile. Both the pump and probe are nominally incident at 45\textdegree~from normal on the sample. The sample consists of 1D nickel gratings of period $P$ = 2000~nm, nominal height $H$ = 12~nm and nominal linewidths $L$ = 200~nm and 500~nm fabricated using electron beam lithography on a diamond substrate, as shown in Fig. \ref{fig:five}(a). The period is set very precisely by the lithography process, while AFM measurements indicate average linewidths of 185~nm and 486~nm for the two gratings and an average height of 11.8~nm, as shown in Fig. \ref{fig:five}(c). The AFM also measures a $\sim$0.7~nm RMS substrate surface roughness. The top $\sim$0.4~nm of the nickel is oxidized as indicated by XRR measurements of a reference sample. The fabrication process produces an amorphous carbon layer on the diamond sample whose thickness varies between $\sim$5~nm underneath the nickel structures and $\sim$20~nm between them, with a density in the range of 1.0-1.8 g/cm$^3$. Details on this layer appear in Ref. \cite{McBennett2024}. 

The blue curve in Fig. \ref{fig:five}(b) shows the static EUV diffraction pattern from the $L$ = 500~nm nickel grating on a CCD camera in a classical diffraction geometry. This one dimensional image is obtained by vertically binning the other dimension (corresponding to the $y$ dimension in section \ref{sec:two}). After excitation by the 800~nm pump pulse, the nickel structures expand and then gradually relax by releasing heat into the substrate. The green curve in Fig. \ref{fig:five}(b) shows the change in counts measured in each pixel on the CCD camera 100~ps after the infrared pump beam excites the sample. We observe an increase in the specular and decrease in the diffracted portion of the EUV probe. The total signal integrated across the $x$ dimension of the camera chip is within the experimental noise at all time delays and for both grating geometries, while transient EUV reflectivity measurements of a uniform nickel film also give poor signal, as shown in Figs. 3 and 4 of the SM \cite{SM}. This suggests that the change in grating reflectivity makes a small contribution to the experimental signal compared to the change in grating height. From this we assume a primarily phase perturbation to the transmittance function.  

Assuming a perfectly square grating and working in scalar diffraction theory, the reconstruction can proceed using Eq.~\ref{eq:gtgreconstruction}, where we have opted to use the analytical expressions for the $\gamma_{\ell}$ given in Eq.~\ref{eq:gtgfouriercomponents} rather than calculate them using RCWA. This is equivalent to the second approach (green curves) in Fig.~\ref{fig:four}(d). On the left-hand side of Eq.~\ref{eq:gtgreconstruction}, the change in diffraction efficiency, $\Delta \tilde{I}_{\ell}(\tau)$, is calculated for each time delay $\tau$ by subtracting consecutive exposures with the pump blocked and unblocked and summing the differential CCD counts across the range of pixels encompassing each diffracted order, before normalizing to the pump blocked exposures. We note that each diffraction order includes the several EUV wavelengths produced by the high harmonic generation process, however, as the photon energies are far away from any electronic resonances, the dynamics encoded in the closely-spaced wavelengths are quite similar and thus the wavelengths are aggregated in the analysis. As $\ell$ increases, the observed change in diffraction efficiency becomes asymmetric between the positive and negative orders, as shown in Fig.~6 of the SM \cite{SM}. This could be caused by a non-uniform initial excitation due to the pump beam's 45\textdegree~incidence, an amplitude perturbation which affects primarily the higher spatial frequencies or the limitations of scalar diffraction theory discussed in section \ref{sec:five}. We assume that the sample perturbation is even with respect to the static unit cell and average $\Delta \tilde{I}_{\ell}(\tau)$ and $\Delta \tilde{I}_{-\ell}(\tau)$ when reconstructing each $\epsilon_{\ell}$. This approach is justified by a comparison of RCWA and Fraunhofer diffraction simulations in SM Fig.~6 \cite{SM}. On the right-hand side of Eq. \ref{eq:gtgreconstruction} , the $\text{sinc}(\ell L/P)$ term can be calculated using the AFM results. The remaining term involving $R_s$, $R_g$ and $\Phi$ is more difficult to calculate, since it depends on the precise EUV wavelength, incidence angle and material properties (see Fig. 5 in the SM \cite{SM}). However, this term is an overall scalar which does not affect the relative magnitudes of the reconstructed $\epsilon_{\ell}(\tau)$. In the present approximation, it is important for determining the size of the perturbation but not its shape.

Figure \ref{fig:five}(d) shows the shape of the reconstructed $\epsilon_{\ell}$ for both gratings as a function of pump-probe delay time. Each curve represents a different $\epsilon_{\ell}$, with the brightest curve showing $\epsilon_1$, obtained from the first diffracted order, and the darkest curve showing the highest $\epsilon_{\ell}$ whose corresponding diffracted order is captured on the CCD camera. In the case of the $L=$ 500~nm $P=$ 2000~nm grating, we do not reconstruct $\epsilon_4$, because the static diffraction efficiency is small, since $4 L / P \sim 1$, and we therefore expect the reconstruction quality to be poor, as illustrated in Fig.~\ref{fig:four}(b). The experimental reconstructions are the average of three scans; each scan consists of multiple sweeps through full time delay range to ensure data repeatability and multiple exposures with the pump blocked and unblocked at each delay. As shown in Fig.~\ref{fig:five}(e), the experimentally reconstructed $\epsilon_{\ell}$ closely resemble the results of a finite element calculation using the model introduced in section \ref{sec:four}. Because the experimental data does not exhibit strong acoustic oscillations, the finite element calculation is performed in the quasi-static approximation discussed in section \ref{sec:four}. The finite element calculation uses the material parameters for nickel and diamond given in Table \ref{tab:one} and represents the effect of the amorphous carbon and oxide layers by including a thermal boundary resistance of 5.5~nK$\cdot$m$^2$/W between the diamond and nickel, which was estimated in a previous work \cite{McBennett2024}. The $\epsilon_{\ell}$ are calculated from the finite element results by converting the vertical surface displacement $u_z(x,\tau)$ to the perturbed transmittance function using Eq.~\ref{eq:phasetoheight} and then Fourier transforming.

The insets in Fig.~\ref{fig:five}(e) show the vertical surface displacement across one unit cell calculated by the finite element model 100~ps after the excitation. The red and purple curves show the displacement after a low pass filter which eliminates all the Fourier components not considered in the experiment, while the gray curves show the unfiltered calculation. The corresponding insets in Fig.~\ref{fig:five}(d) show the vertical surface displacement calculated from the reconstructed $\epsilon_{\ell}$ using Eq.~\ref{eq:phasetoheight}. It is likely that the sidewalls of the grating expand horizontally during heating and that there are edge effects as discussed in section \ref{sec:five}, which should be considered in a more rigorous treatment. However, the qualitative agreement is quite good, suggesting that with adequate knowledge of the static profile it is possible to spatially reconstruct a small dynamic phase perturbation to a periodic sample using only a scatterometry measurement.

\section{Discussion}
\label{sec:discussion}

The dynamic imaging of periodic structures (DIPS) method provides a promising route to spatially reconstruct dynamics \textit{within} a unit cell of a periodic system using the changing diffraction efficiencies in dynamic scatterometry data. In particular, it takes advantage of interference between the fields generated by the static profile and dynamic perturbation and exploits a case where off-diagonal terms in a matrix relating the perturbation to the change in far field diffraction disappear. This diagonal case, given in Eq.~\ref{eq:linearphasediagonal}, is appealing, since it allows each Fourier component of the dynamic perturbation to be treated separately rather than be solved via a system of linear equations. In most experiments, precise knowledge of the static profile is limited to the first few Fourier components, allowing for a straightforward propagation of uncertainty between each static $\gamma_{\ell}$ and the corresponding dynamic $\epsilon_{\ell}$. 

In the previous sections, we demonstrated that the reconstruction of experimental data works best in situations where there is a pure even phase perturbation to the transmittance function; in the case of a simultaneous amplitude perturbation, one must resort to approximate methods which invoke physical knowledge of the system. While the reconstruction of the experimental data agrees qualitatively with the modeling, we note that the experimental setup used in this work is not optimized for the DIPS method. To increase the utility of these reconstructions, one could design an experiment in a transmission geometry, including a normal incidence or conical diffraction geometry for the probe, a minimal sample-to-camera distance to capture many diffraction orders (\textit{i.e.}, a high numerical aperture), carefully chosen materials such that $R_s/R_g\sim1$, and $L,P$ combinations that minimize the error in Fig.~\ref{fig:four}. Optimization of these design parameters would allow for reconstruction of the maximum number of $\epsilon_l$ possible with minimized induced errors. However, even in the optimized geometry, the limitations of the paraxial and Kirchhoff approximations still exist. Using additional wavelengths and incidence angles, as in static scatterometry, could be beneficial to this technique; however, further work is necessary to show this utility.

Despite the effectiveness of this approach as shown in section~\ref{sec:experiment}, there exist challenges in general implementation to arbitrary samples and geometry. While the static field can be calculated from knowledge of the sample (\textit{i.e.}, from AFM characterization), this approach is susceptible to errors stemming from incomplete knowledge of the sample (\textit{e.g.}, densities, oxide layers, uncertainty in scattering factors).  A promising avenue to precisely determine the field from the static profile is to use a (relatively slow) coherent diffraction imaging technique like ptychography to solve for the static field \cite{Wang2023,Esashi2023}, and then use DIPS to record the dynamics by analyzing individual diffraction patterns at each time step. As shown in Fig. 1 of the SM \cite{SM}, this is even a valid approach in cases where the paraxial approximation does not strictly hold; the primary complication arises when one wishes to extract the physical dynamics in the sample that caused the (somewhat simplified) dynamics in the field leaving the sample. 

Nonetheless, we propose and demonstrate in section~\ref{sec:experiment} that the DIPS method can be used on experimental data to track nanoscale energy transport ({\it e.g.} heat flow) with ultrafast time resolution. Recent work by Karl Jr. {\it et al.} used ptychography with ultrafast EUV sources to reconstruct a time- and space-resolved movie of a laser-excited isolated Ni nanoantenna on silicon achieving sub-$100$ nm lateral resolution for 14 different time delays ~\cite{Karl2018}. For this fully-general method, 82 2D scatter patterns are needed to reconstruct a single pump unblocked image which, in total, required $14.5$ hours of acquisition time and $>$15GB of data. In comparison, the measurements discussed in this work only require $<$10 minutes for over 150 different time delays and, when used in conjunction with the DIPS method, still capture spatial information assuming the sample is periodic and well-characterized. We note that these two approaches are not directly comparable as the approach by Karl Jr. {\it et al.} captures 2D spatial images on the sample's time-dependent complex reflectance without prior knowledge of the sample, while the DIPS only reconstructs a 1D waveform on a single unit cell in a 1D periodic structure with prior knowledge of the sample under specific conditions. However, we note that recent works acquired dynamic EUV scatterometry data on the cooling of 2D periodic nanoscale heaters \cite{Beardo2021} which could, in theory, also be used with DIPS; however, the extension of the DIPS method to 2D periodic samples is the subject of future work. While the DIPS method can provide 1D images in real space, we note that in some cases comparison with theory could, and may more naturally, be done in the spatial frequency domain, specifically, at the $\epsilon_l(\tau)$ level. For example, new theoretical approaches in understanding nanoscale heat transport often perform calculations in the spatial frequency domain \cite{MinnichSpatialFrequency_2018,SchellingThermalResponse2025} which is well-matched to experimental data used with the DIPS method. It may be possible to validate physical models even at the level of the spatial information encoded in the two-dimensional electric field, especially in nanoscale spin, charge and heat transport where models may predict unique microscopic dynamics. Regardless, this method can provide additional information from the same data set that has not been explored in previous works \cite{Hoogeboom-Pot2016,Frazer2019,Beardo2021}. Furthermore, if the approach can be refined to handle general amplitude perturbations, it could be applied in situations where amplitude perturbations are unavoidable, such as spin dynamics in nanostructured magnets \cite{Shaw2009} and provide a fast complementary approach to synchrotron imaging methods \cite{Donnelly2020}. Alternatively, it could be used to rapidly characterize wafer-scale variations in a \textit{static} parameter, as long as the variation is small relative to the average value.

\newpage

\section*{Acknowledgments}

The authors acknowledge support from the STROBE National Science Foundation Science \& Technology Center, Grant No. DMR-1548924. We also acknowledge support from the National Science Foundation under Grant No. DMR-2117903. Development of the DIPS method was supported by funding from the CHIPS Metrology Program, part of CHIPS for America, National Institute of Standards and Technology, U.S. Department of Commerce. Certain equipment, instruments, software, or materials are identified in this paper in order to specify the experimental procedure adequately.  Such identification is not intended to imply recommendation or endorsement of any product or service by NIST, nor is it intended to imply that the materials or equipment identified are necessarily the best available for the purpose.

\section*{References}

\bibliographystyle{ieeetr}
\bibliography{main}

\end{document}


\title{Supplemental Material: Dynamic Imaging of Periodic Structures using Extreme Ultraviolet Scatterometry}

\maketitle

\section{Non-paraxial effects on the derivation}

\begin{figure}[H]
    \centering
    \includegraphics[width=1\linewidth]{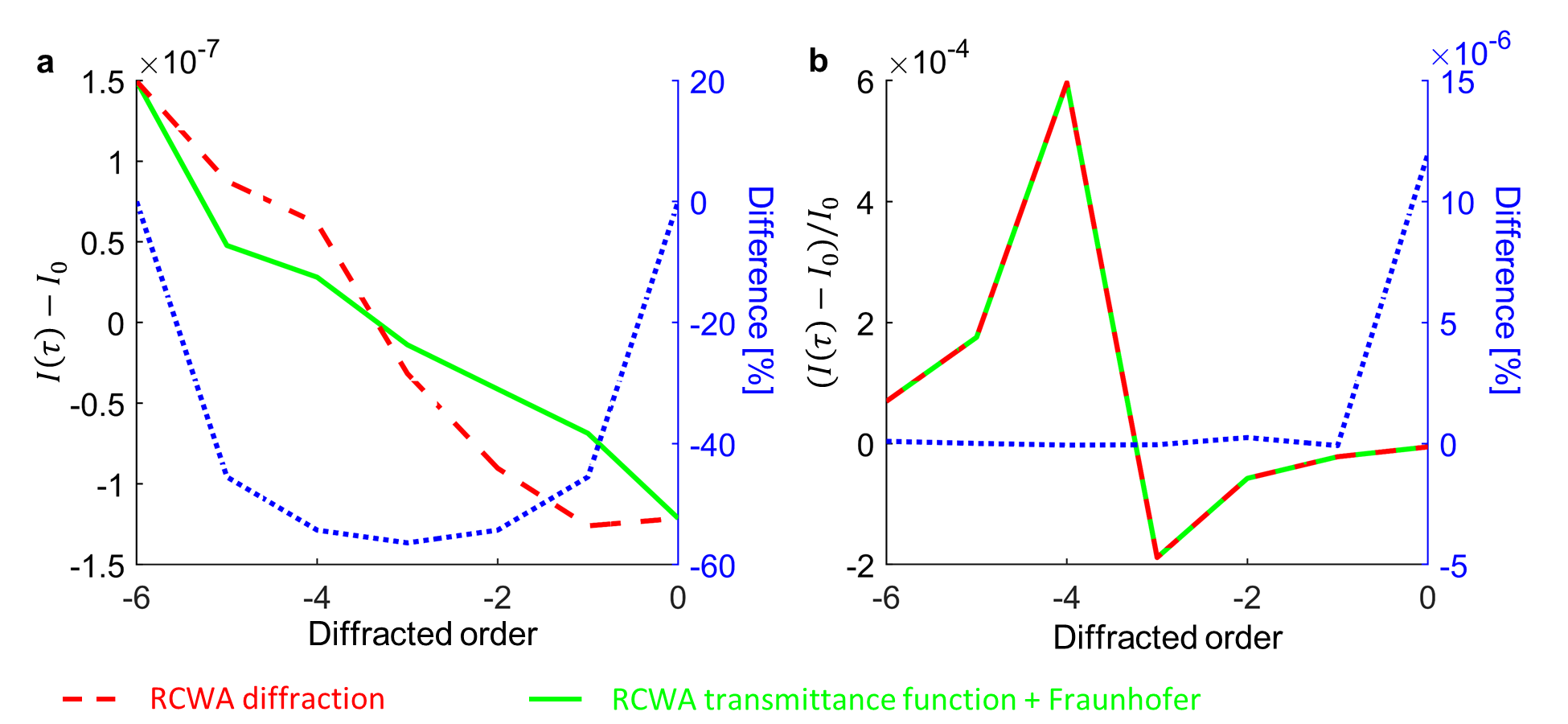}
    \caption{\textbf{Normalized change in diffracted intensity outside the paraxial approximation} (a) Change in diffracted intensity $I(\tau) - I_0$ versus diffracted order for 30~nm light incident at 64 degrees from normal on a square nickel grating with $H$ = 50~nm, $L$ = 30~nm and $P$ = 100~nm on a silicon substrate after a random vertical surface displacement. The red dashed line is a full RCWA calculation while the solid green line uses Fraunhofer (paraxial) diffraction of the object plane electric field calculated using RCWA. The dashed blue line (second axis) gives the percent difference. (b) The same calculation after normalizing to the static diffraction efficiencies.}
    \label{fig:one}
\end{figure}

To show numerically that the change in diffracted intensity in each order is unaffected by the paraxial approximation, we generate random vertical displacements to a square nickel grating geometry on a silicon substrate with linewidth $L=30$~nm, period $P=100$~nm and height $H=50$~nm using Eq.~20 in the main text. Next, the far field diffraction efficiencies for the static and perturbed grating are calculated outside the paraxial approximation using the RCWA code described in section V of the main text \cite{reticolo}, for 30~nm light incident at 64 degrees from normal. Fig.~\ref{fig:one} (red dashed curve) shows the non-normalized and normalized RCWA change in diffraction efficiency in each order due to the perturbation. In order to compare to an equivalent paraxial diffraction simulation, the RCWA code next calculates the reflected electric field for the static and perturbed geometries several hundred nanometers above the grating surface, where evanescent modes are negligible, and then reverse propagates it using the angular spectrum method \cite{Goodman2005} to the plane defined by the top of the grating. From here, a Fourier transform determines the paraxial (Fraunhofer) diffraction efficiency in the far field. While the non-normalized change in Fraunhofer diffraction efficiency differs from the RCWA calculation, as shown in Fig.~\ref{fig:one}(a) (green curve), the normalized value in Fig.~\ref{fig:one}(b) is the same. This is true regardless of perturbation size, incidence angle and structure aspect ratio. We speculate that this might have direct implications in the validity of certain pre-processing steps common in reflection-mode ptychography experiments; specifically, it might indicate some rigor behind the step of dividing the diffraction patterns by an intensity distribution that varies over the detector.

\section{Example random vertical surface displacement for stochastic simulations}

\begin{figure}[H]
    \centering
    \includegraphics[width=1\linewidth]{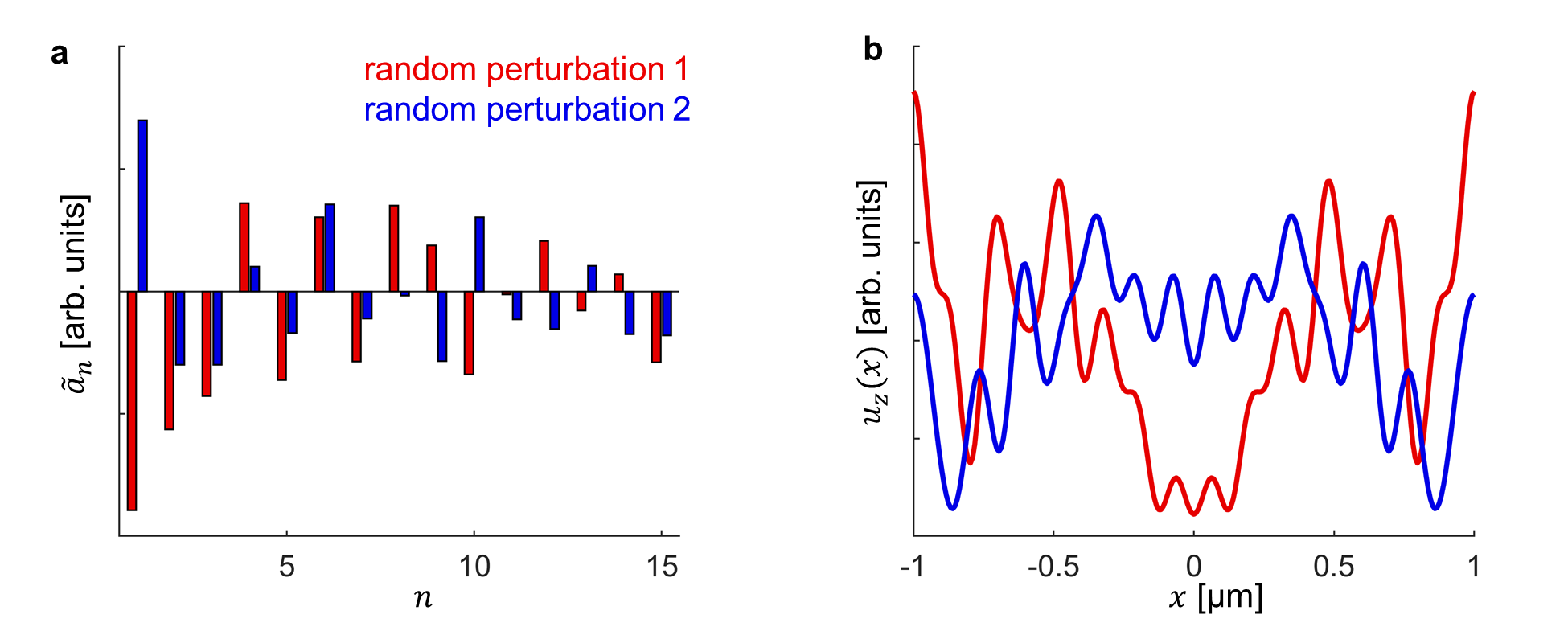}
    \caption{\textbf{Example random vertical surface displacements used for simulations in section V of the main text.} (a) Fourier coefficients $\tilde{a}_n$ drawn from the uniform distributions [$-a_n$,$a_n$], with $a_n$ = $a_1 n^{-1/2}$. (b) Real space vertical surface displacement $u_z(x)$ calculated using Eq. 20 of the main text, $u_z(x) = \sum_{n=1}^N \tilde{a}_n \cos(2 \pi x n /P)$.}
    \label{fig:two}
\end{figure}

\section{Phase-only perturbation in dynamic EUV scatterometry experiment}

\begin{figure}[H]
    \centering
    \includegraphics[width=1\linewidth]{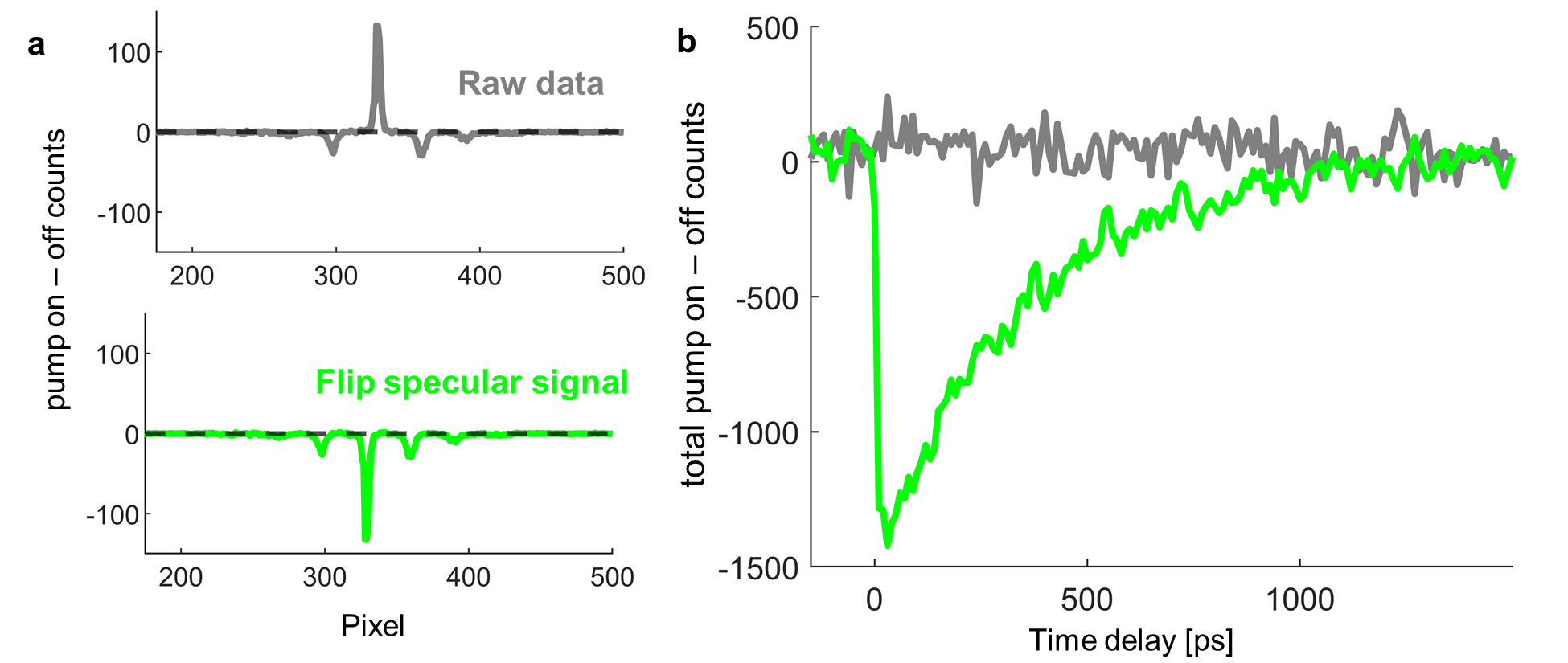}
    \caption{\textbf{Phase-only perturbation in dynamic EUV scatterometry experiment} (a) Pixel-wise difference in CCD counts with the pump blocked and unblocked for the L = 500~nm grating at 100~ps delay: raw data (gray) and data with specular signal flipped (green). (b) Integrated difference in CCD counts as a function of pump-probe time delay. The raw data shows no signal when integrated across all pixels, indicating that, at a minimum, Re($\epsilon_0$) $\approx$ 0.}
    \label{fig:three}
\end{figure}

\begin{figure}[H]
    \centering
    \includegraphics[width=1\linewidth]{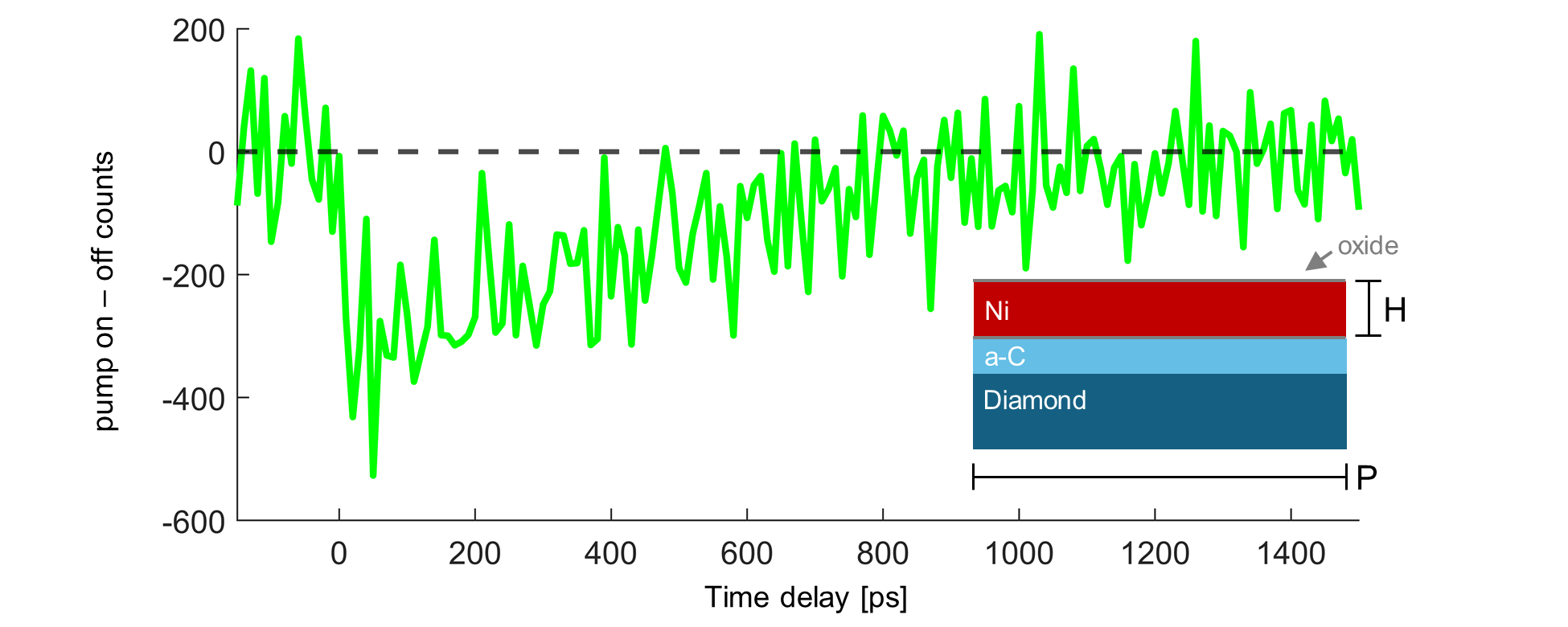}
    \caption{\textbf{Transient EUV reflectivity signal on a uniform nickel film.} The transient reflectivity data here was taken with $\sim$20\% fewer incident EUV photons per frame, and an infrared pump fluence of 28 mJ/cm$^2$ as compared to 27 mJ/cm$^2$ for the scatterometry data in Fig. \ref{fig:three}.}
    \label{fig:four}
\end{figure}

\section{Estimating the perturbation magnitude}

The reconstruction in section VI of the main text assumes a phase perturbation which is even with respect to a square grating of height $H$, linewidth $L$, period $P$ and grating and substrate reflectivities $r_g = R_g e^{i \phi_g}$ and $r_s = R_g e^{i \phi_g}$, respectively. In the Kirchhoff approximation, the $\epsilon_{\ell}$ Fourier components of the perturbation can be obtained from

\begin{equation} \label{eq:SI1}
	\Delta \tilde{I}_{\ell}(\tau) = i \epsilon_{\ell}(\tau) \frac{2P}{L} \left[ \text{sinc}\left( \frac{\ell L}{P}\right) \right ]^{-1} \frac{R_gR_s\sin(\Phi)}{R_g^2 + R_s^2 - 2 R_g R_s \cos(\Phi)}.
\end{equation}

In Eq. \ref{eq:SI1}, $\Phi = \phi_s+\phi_H-\phi_g$, where $\phi_H = 4\pi H \cos{\theta} / \lambda$ is the geometric phase imparted by the grating for light of wavelength $\lambda$ incident at an angle $\theta$ from normal. As discussed in the main text, the normalized change in diffraction efficiency $\Delta \tilde{I}_{\ell}(\tau)$ can be calculated directly from the experimental data, while the $\text{sinc}\left( \ell L / P \right)$ term can be obtained from AFM measurements. This is sufficient to determine the shape of the reconstructed perturbation; however, to determine its magnitude it is also necessary to know the value of the fraction on the right-hand side of Eq.~\ref{eq:SI1} involving $R_g$, $R_s$ and $\Phi$. 

It is possible to estimate this coefficient in Eq.~\ref{eq:SI1} by using the EUV wavelength, sample geometry and material densities to directly calculate $R_s$, $R_g$ and $\Phi$ from elemental scattering factors, material density and the Parratt formalism, as discussed in section IV of the main text. The high harmonic generation process used in the present EUV scatterometry experiment generates a comb consisting primarily of three higher harmonics of the fundamental $\sim$800~nm wavelength spaced by several nanometers, which overlap on the CCD camera for the lower diffracted orders (see main text Fig. 5). The wavelengths can be estimated using the grating equation,

\begin{equation} \label{eq:SIgtg}
	\sin \theta_i + \sin \theta_m = \frac{m\lambda}{P},
\end{equation}

where $P = $2000~nm is the grating period, $\theta_i = $ 45\textdegree~is the angle of incidence and $\theta_m$ is the angle of diffraction for order $m$. The diffraction angle can be calculated by carefully measuring the locations of the specular beam and diffracted orders on the CCD camera (26~\textmu m pixel size) and the sample to camera distance, which we estimate to be 37 $\pm$ 1~mm. This corresponds to wavelengths in the range of 28-31~nm, 30-33~nm and 34-36~nm for the three main EUV harmonics, as expected for high harmonic generation in argon. The EUV light is therefore far from the carbon and nickel resonances and sensitive primarily to the vertical surface displacement and material density. However, because of uncertainty in the density and thickness of the amorphous carbon interlayer, as discussed in Ref. \cite{McBennett2024}, it is very difficult to directly calculate $R_s$, $R_g$ and $\Phi$ using this approach.

Alternatively, the coefficient in Eq.~\ref{eq:SI1} can be estimated using measurements of the static reflectivity and diffraction efficiencies. The relative magnitude of $R_s$ and $R_g$ is first approximated by measuring the EUV reflectivity of the substrate as compared to a large nickel film patterned adjacent to the gratings. For the present sample this measurement indicates $R_s^2 \approx 2.4 R_g^2$. To estimate $\sin(\Phi)$ and $\cos(\Phi)$, which involve the material and geometric phases, we consider the ratio of the static diffraction efficiency in the $\ell$th diffracted order relative to the specular beam, which can be seen in Eq. 7 of the main text to be $|\gamma_\ell|^2 / |\gamma_0|^2$. Using the expressions in Eq. 14 of the main text for a square grating in the Kirchhoff approximation, this ratio can be written as

\begin{equation} \label{eq:SI2}
	\frac{|\gamma_\ell|^2}{|\gamma_0|^2} = \frac{\frac{L^2}{P^2} \text{sinc}^2(\frac{\ell L}{P})(R_g^2 + R_s^2 - 2R_gR_s \cos(\Phi))}{\frac{L^2}{P^2} R_g^2 + R_s^2 (1-\frac{L}{P})^2 + 2\frac{L}{P}(1-\frac{L}{P})R_g R_s \cos(\Phi)}.
\end{equation}

Using the estimate of $R_s^2 \approx 2.4 R_g^2$ described above, and experimental measurements of $|\gamma_{\ell}|^2 / |\gamma_0|^2$ in the frames with the infrared pump blocked, we solve Eq.~\ref{eq:SI2} for $\cos(\Phi)$. As shown in Fig. \ref{fig:four}, the results vary depending on the diffracted order and grating, but cluster around $\cos(\Phi) \approx 0$. This constrains the value of $\sin(\Phi)$ to be approximately $\pm 1$. To determine the correct sign we consider the change in diffraction efficiency in Fig. 5(b) of the main text. To first order, this change in diffraction efficiency is caused by an increase in nickel structure height $\Delta H$ after it absorbs the 800~nm pump pulse. The $\epsilon_{\ell}$ Fourier components associated with a uniform $\Delta H$ change in structure height are

\begin{equation} \label{eq:SI3}
	\epsilon_{\ell} = -\frac{4 \pi i \cos \theta}{\lambda} \frac{L}{P} \cdot \text{sinc}\left( \frac{\ell L}{P}\right)\Delta H,
\end{equation}

where $\theta$ is the EUV incidence and $\lambda$ is the wavelength. Inserting this into Eq.~\ref{eq:SI1} and noting that $R_g^2 + R_s^2 - 2R_gR_s \cos(\Phi)$ is always positive, it is apparent that $\sin(\Phi)$ must be negative to produce the appropriate experimental diffraction profile. Inserting $\cos(\Phi) = 0$, $\sin(\Phi) = -1$ and $R_s^2 = 2.4 R_g^2$ into Eq.~\ref{eq:SI1} results in reconstructed $\epsilon_{1}$ of order 10$^{-3}$ at around 100~ps after the excitation for the $L =$ 500~nm grating, corresponding to a height change on the order of 0.1 angstrom. For nickel, which has a thermal expansion coefficient of $12.8 \times 10^{-6}$  1/K \cite{Kollie1977}, this suggests a physically reasonable $\sim$100~K temperature rise.

\begin{figure}
    \centering
    \includegraphics[width=1\linewidth]{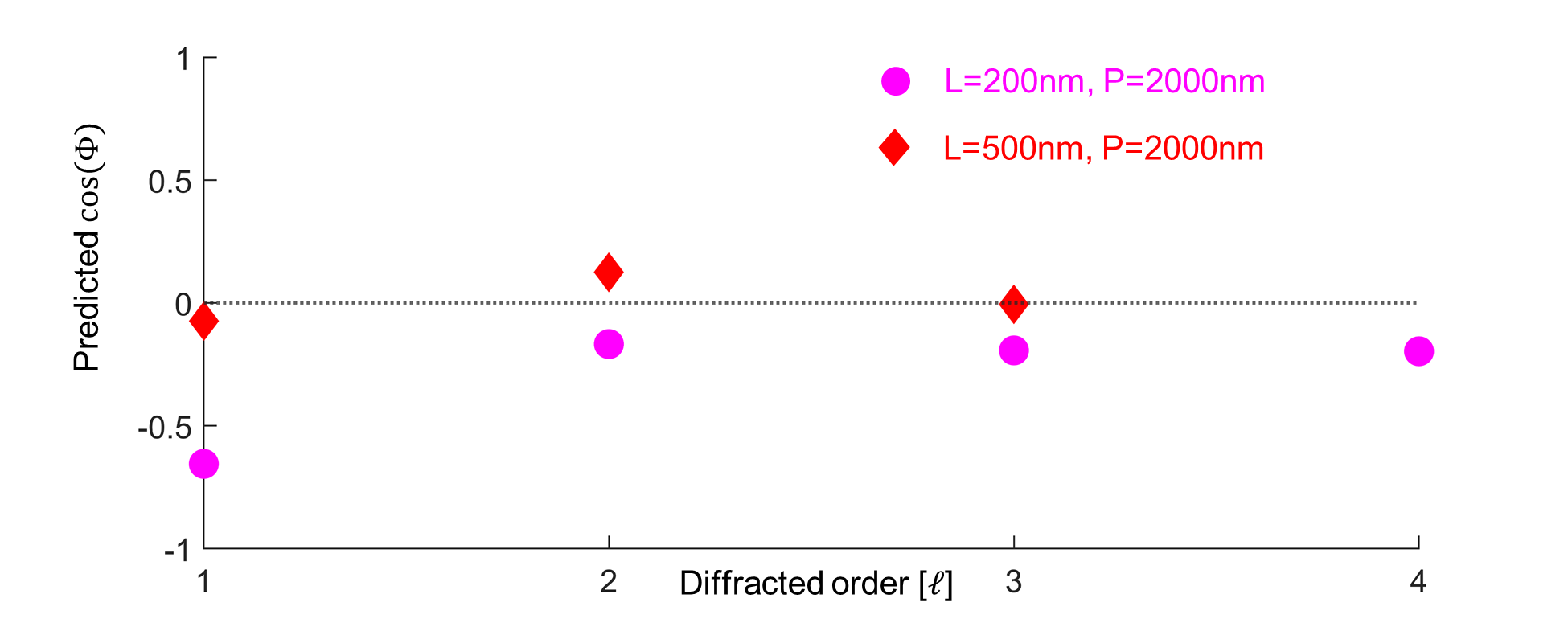}
    \caption{\textbf{Estimate of $\cos(\Phi)$ from the static diffraction efficiencies}. The calculation in Eq. \ref{eq:SI2} is repeated using the average of each positive and negative diffracted order ($\pm \ell$) for both grating geometries.}
    \label{fig:five}
\end{figure}

\section{Simulations of asymmetrical experimental diffraction profiles}

\begin{figure}[H]
    \centering
   \includegraphics[width=1\linewidth]{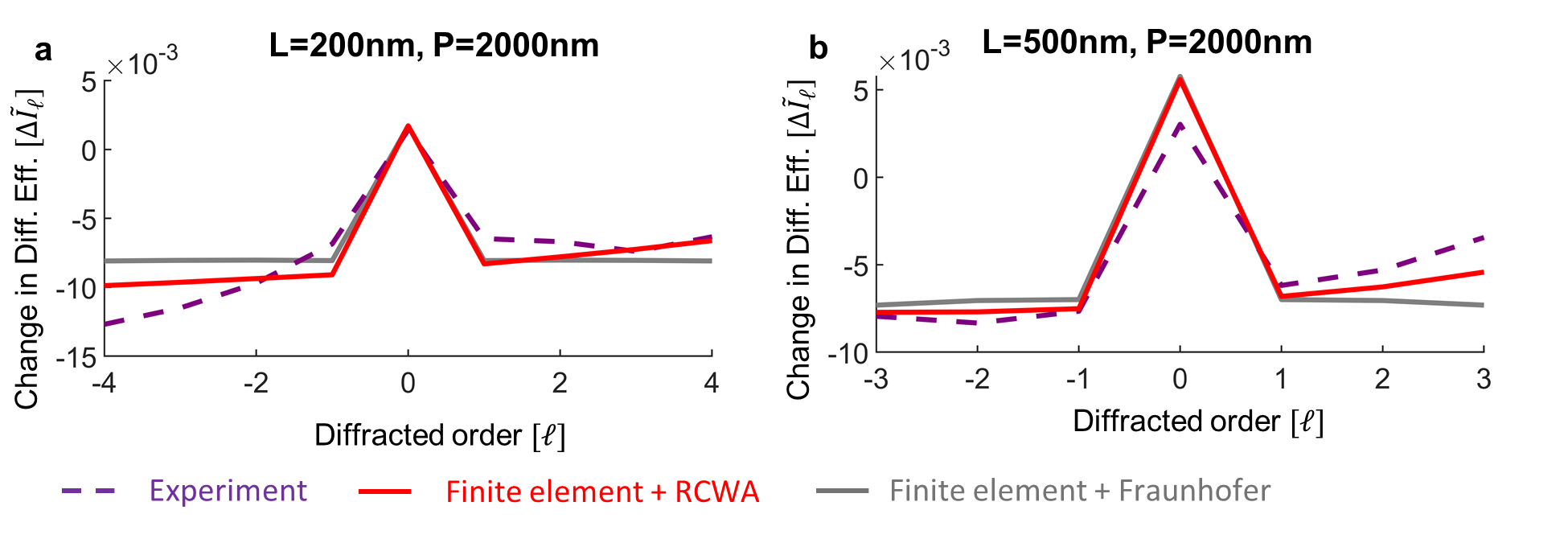}
    \caption{\textbf{Change in diffraction efficiency $\Delta \tilde{I}_{\ell}$ vs. diffracted order $\ell$}. (a) Results for the $L=$ 200~nm $P=$ 2000~nm grating. The dashed purple curve shows the dynamic EUV scatterometry experiment averaged over all scans and time delays above 20~ps. The solid curves show the results of the finite element simulation of the vertical surface displacement at 100~ps time delay. The red curve is an RCWA simulation of the change in diffraction efficiency due to the displacement, while the gray curve uses Fraunhofer diffraction, with Eq. 16 of the main text used to convert the displacement into a perturbed transmittance function. (b) Equivalent calculation for the $L=$ 500~nm $P=$ 2000~nm grating. The average of $\Delta \tilde{I}_{+\ell}$ and $\Delta \tilde{I}_{-\ell}$ in the RCWA simulation gives approximately the Fraunhofer result, supporting the decision to average the plus and minus diffracted orders when reconstructing each $\epsilon_{\ell}$ from the experimental data. }
    \label{fig:six}
\end{figure}

\section*{References}

\bibliographystyle{ieeetr}
\bibliography{sm}
\nocite{*}